\pdfoutput=1
\documentclass[journal]{IEEEtran}
\usepackage[numbers,sort&compress]{natbib}
\usepackage{amsmath}
\usepackage{graphicx}
\usepackage[hyphens]{url}
\usepackage{breakurl}
\usepackage{makecell}
\usepackage{bm}
\usepackage{booktabs}
\usepackage{color}
\usepackage{multirow}
\usepackage{amsthm,amssymb}
\usepackage{mathrsfs}
\ifCLASSINFOpdf
\else
\fi

\begin{document}
\bstctlcite{IEEEexample:BSTcontrol}
%
\title{A Four-Stage Data Augmentation Approach to ResNet-Conformer Based Acoustic Modeling for Sound Event Localization and Detection}
%
%
%

\author{Qing Wang, Jun Du, Hua-Xin Wu, Jia Pan, Feng Ma,
        and Chin-Hui Lee,~\IEEEmembership{Fellow,~IEEE}
\thanks{Q. Wang and J. Du are with University of Science and Technology of China, Hefei 230027, China (e-mail: qingwang2@ustc.edu.cn, jundu@ustc.edu.cn).}
\thanks{H.-X. Wu, J. Pan, and F. Ma are with iFlytek, Hefei 230088, China
(email: hxwu2@iflytek.com, jiapan@mail.ustc.edu.cn, fengma@iflytek.com).}
\thanks{C.-H. Lee is with the School of Electrical and Computer Engineering,
Georgia Institute of Technology, Atlanta, GA 30332-0250 USA (e-mail: chl@ece.gatech.edu).}}

%
%

\markboth{IEEE/ACM TRANSACTIONS ON AUDIO, SPEECH, AND LANGUAGE PROCESSING}
{Shell \MakeLowercase{\textit{et al.}}: Bare Demo of IEEEtran.cls for IEEE Journals}
%

\maketitle

\begin{abstract}
In this paper, we propose a novel four-stage data augmentation approach to ResNet-Conformer based acoustic modeling for sound event localization and detection (SELD). First, we explore two spatial augmentation techniques, namely audio channel swapping (ACS) and multi-channel simulation (MCS), to deal with data sparsity in SELD. ACS and MDS focus on augmenting the limited training data with expanding direction of arrival (DOA) representations such that the acoustic models trained with the augmented data are robust to localization variations of acoustic sources. Next, time-domain mixing (TDM) and time-frequency masking (TFM) are also investigated to deal with overlapping sound events and data diversity. Finally, ACS, MCS, TDM and TFM are combined in a step-by-step manner to form an effective four-stage data augmentation scheme. Tested on the Detection and Classification of Acoustic Scenes and Events (DCASE) 2020 data set, our proposed augmentation approach greatly improves the system performance, ranking our submitted system in the first place in the SELD task of the DCASE 2020 Challenge. Furthermore, we employ a ResNet-Conformer architecture to model both global and local context dependencies of an audio sequence
and win the first place in the DCASE 2022 SELD evaluations.
\end{abstract}

\begin{IEEEkeywords}
Spatial data augmentation, sound event detection, sound source localization, direction of arrival, Conformer.
\end{IEEEkeywords}

%
\IEEEpeerreviewmaketitle

\section{Introduction}
\IEEEPARstart{S}{ound} event localization and detection (SELD) is a task to detect the presence of individual sound events and localize their arriving directions. Humans can correctly identify and localize multiple sound events overlapping both temporally and spatially in an audio signal, but it is very challenging for machines. Effective SELD is of great importance in many applications. For instance, SELD-enabled robots can perform search and rescue missions when detecting the presence of a fire, an alarm, or a scream, and localizing them. In teleconferences, an active speaker can be recognized and tracked, making it possible to use beamforming techniques to enhance speech and to improve automatic speech recognition (ASR) \cite{wang1997voice,swietojanski2014convolutional}. Intelligent homes and smart cities can also employ SELD for acoustic scene analysis and audio surveillance \cite{valenzise2007scream,foggia2015audio}.

The SELD problem consists of two key issues denoted as sound event detection (SED) and sound source localization (SSL). SED aims to recognize individual sound events in an audio sequence together with their onset and offset times. Early SED methods \cite{heittola2010audio,mesaros2010acoustic,heittola2013context} came from the ASR field, with Gaussian mixture models (GMM) and hidden Markov models (HMM) used for acoustic modeling. However, when overlapping events occur, the detection results were often unsatisfactory. Non-negative matrix factorization (NMF) based algorithms were used to learn a dictionary of basis vectors and then separate sound sources 
\cite{gemmeke2013exemplar,mesaros2015sound}. Nevertheless, this method is not robust in noisy environments.

Recently, deep neural network (DNN) architectures in various forms have been successfully employed for SED. Feed-forward neural networks (FNN) were used for sound event classification, which greatly outperformed support vector machines (SVM) at low signal-to-noise ratio (SNR) levels \cite{mcloughlin2015robust}. Convolutional neural networks (CNN) \cite{piczak2015environmental,zhang2015robust,phan2016robust} and recurrent neural networks (RNN) \cite{wang2016audio,parascandolo2016recurrent,hayashi2017duration} were also adopted for SED. Capsule neural networks (CapsNet) \cite{sabour2017dynamic} which were originally proposed for image classification, have been used to separate individual sound events from an overlapping mixture by selecting the most representative spectral features of each sound event \cite{vesperini2019polyphonic,liu2018capsule}. State-of-the-art results for SED were achieved by a convolutional recurrent neural network (CRNN) \cite{cakir2017convolutional,adavanne2018sound,cao2019polyphonic}, a recently published architecture that combined CNN, RNN, and FNN.

SSL aims to estimate the direction-of-arrival (DOA) for each sound source. Various algorithms have been proposed for DOA estimation. These approaches can be categorized into two kinds: conventional and DNN-based. Conventional DOA estimation approaches include multiple signal classification (MUSIC) \cite{schmidt1986multiple}, estimation of signal parameters via rotational invariance technique (ESPRIT) \cite{roy1989esprit,teutsch2006acoustic}, and steered response power phase transform (SRP-PHAT) \cite{knapp1976generalized,brandstein1997robust,do2007real}, which rely on a sound field model. DNN-based approaches, however, do not rely on preassumptions a priori assumptions about array geometries and have superior generalization ability to unseen scenarios because of their high regression capability \cite{ferguson2018sound,liu2018direction,adavanne2018direction}. Previous studies have proposed a DOA estimation method for overlapping sources by combining sound intensity vector methods \cite{pavlidi20153d,hafezi2017augmented} and DNN-based separation \cite{yasuda2020sound}. A DNN-based phase difference enhancement for DOA estimation was proposed in \cite{pak2019sound}, showing better results than direct regression to DOA representation.

For supervised methods, one key factor that affects the system performance is the size of the training data. It is difficult for DNNs to learn the relationship between the input and output for small datasets. And models trained with small datasets tend to face overfitting problem. In machine learning, data augmentation is an effective strategy to overcome a lack of training data and to alleviate overfitting, which has been widely used in many areas, such as ASR, sound classification, image classification, and computer vision \cite{cui2015data,salamon2017deep,perez2017effectiveness,simard2003best}. In this study, we focus on data augmentation techniques to deal with limited training data in DNN-based acoustic modeling for SELD. The configuration of the microphone array is an important factor that affects the performance of a SELD system. Several kinds of microphone arrays are used in the literature, such as spherical array \cite{politis2020dataset}, mobile phone \cite{pertila2021mobile}, circular array \cite{brousmiche2020secl}, and wearable array \cite{nagatomo2022wearable}. Indeed, most related researches use spherical microphone arrays to collect experimental data \cite{evers2020locata,politis2021dataset,guizzo2022l3das22}. Due to the difficulty of recording and annotation procedure for sound scene data, these SELD datasets are usually of small size and not suitable for training DNN-based methods. This motivates us to exploit the spatial augmentation approach for SELD.

For the SED task, time stretching, pitch shifting, equalized mixture data augmentation (EMDA) \cite{takahashi2017aenet}, and mixup \cite{zhang2017mixup} are effective to improve system performances \cite{parascandolo2016recurrent,lu2017bidirectional,takahashi2016deep,shimada2020sound}. Since these augmentation techniques increase the diversity of input audio signals, they are also applicable to the SELD task. In addition to SED, the SELD task addresses SSL, i.e., DOA estimation by utilizing DOA labels of single or multiple sound sources. It should be noted that the spatial information may be affected when the audio signals are modified by data augmentation methods. To the best of our knowledge, there exists only a few data augmentation studies for DOA estimation. He \emph{et al.} proposed to mix single-source segments for multi-speaker DOA estimation \cite{he2021neural}. Mazzon's team first proposed a spatial augmentation method based on properties of first-order Ambisonics (FOA) sound encoding \cite{mazzon2019first}. It focused on expanding the representation of the DOA subspace for the FOA data set and was effective to reduce DOA errors. However, the authors did not investigate its effectiveness to deal with overlapping sound events, which is necessary for future SELD applications in real-life acoustic scenes with potentially overlapping events.

In this study, we investigate a number of novel approaches to spatial data augmentation for acoustic modeling in SELD. We first propose two techniques, namely audio channel swapping (ACS) and multi-channel simulation (MCS), to increase the diversity of DOA representations from the limited training data. ACS performs transformation on audio channels, which is based on the physical and rotational properties of the spherical microphone array. The MCS approach aims to simulate new multi-channel data by estimating spatial information carried by audio segments containing isolated and static sound events. A complex Gaussian mixture model (CGMM) \cite{higuchi2016robust} is used to estimate time-frequency (T-F) masks and a generalized eigenvalue (GEV) beamformer \cite{warsitz2007blind} is employed to obtain enhanced spectra which are combined with spatial information to simulate multi-channel data. In addition to ACS and MCS, we adopt two other augmentation techniques for SELD, namely time-domain mixing (TDM) which randomly mixes two individual sound events in the time domain and is similar to EMDA \cite{takahashi2017aenet} and time-frequency masking (TFM) which randomly drops several consecutive frames or frequency bins of spectra features \cite{park2019specaugment,zhang2019data}.

By combining these four complementary techniques in a stage-by-stage manner without the Conformer, our submitted ResNet-GRU based system \cite{Du2020_task3_report} achieved the best performance for the SELD task of the Detection and Classification of Acoustic Scenes and Events (DCASE) 2020 Challenge \cite{DCASE2020_task3_report}. To further improve acoustic modeling in this study, we also adopt a Conformer that combines convolution and transformer and achieves state-of-the-art results in ASR \cite{gulati2020conformer}. The Conformer module is a novel combination of self-attention and convolution, with self-attention capturing global dependencies and convolution learning local features in an audio sequence. We incorporate the Conformer framework into the ResNet used in our DCASE 2020 system and evaluate it in real spatial sound scenes. The resulting ResNet-Conformer architecture \cite{Du_NERCSLIP_task3_report} ranks the first place in the SELD task of the DCASE 2022 Challenge \cite{DCASE2022_task3_report}.

Our major contributions can be summarized as follows:

1) presenting a novel ACS spatial augmentation method to expand data sets of the DCASE 2020 and 2022 Challenges based on symmetrical distribution characteristics of spherical microphone array;

2) proposing a novel MCS spatial augmentation technique to increase the number of DOA representations for static sound events by estimating spatial information carried by audio signals;

3) incorporating a Conformer module into a ResNet system to form a ResNet-Conformer architecture that captures both global and local context dependencies in an audio sequence;

4) designing a set of comprehensive experiments for the DACSE 2020 SELD task to show the effectiveness of the proposed four-stage data augmentation approach to acoustic modeling for the proposed ResNet-Conformer architecture.

The remainder of the paper is organized as follows. Section~\ref{sec:DA} describes the spatial data augmentation approaches, especially for ACS and MCS. Section~\ref{sec:conformer} details the Conformer architecture. Experimental results and analysis are presented in Section~\ref{sec:exp}. Finally we conclude the paper in Section~\ref{sec:conc}.

\section{Four-Stage Data Augmentation}
\label{sec:DA}

\subsection{Audio Channel Swapping (ACS)}
\label{subsec:ACS}

The ACS spatial augmentation method is proposed to deal with small SELD datasets collected by spherical microphone arrays, which play an important role in sound field analysis and have gained a lot of popularity in recent years. The proposed ACS method could be applied to two kinds of data formats. The first is first-order Ambisonics (FOA) format which can be obtained by directly using Ambisonics microphones \cite{kurz2015comparison}, e.g., SoundField ST450, SoundField SPS200, Oktava MK4012 or by converting the spherical microphone array signals to first-order Ambisonics with measurement-based encoding filters \cite{politis2017comparing}. The second data format is tetrahedral microphone array (MIC) in which the placement of the four microphones on the surface of the sphere needs to meet certain requirements. The azimuth angles of four microphones must be exactly the same as described in the next paragraph, whereas there is no explicit requirement for the elevation angles, as long as the distances from the four microphones to the xy : z = 0 plane are the same. Generally speaking, the proposed ACS spatial augmentation method is applicable to a wide range of datasets, e.g., data used in the SELD task in DCASE 2019-2022 Challenges \cite{Adavanne2019_DCASE,politis2020dataset,politis2021dataset,politis2022starss22}, in the 3D SELD task in L3DAS21 and L3DAS22 Challenges \cite{guizzo2021l3das21,guizzo2022l3das22}, and the LOCATA Challenge corpus \cite{evers2020locata}.

\begin{figure}[t]
  \centering
  \centerline{\includegraphics[width=0.8\columnwidth]{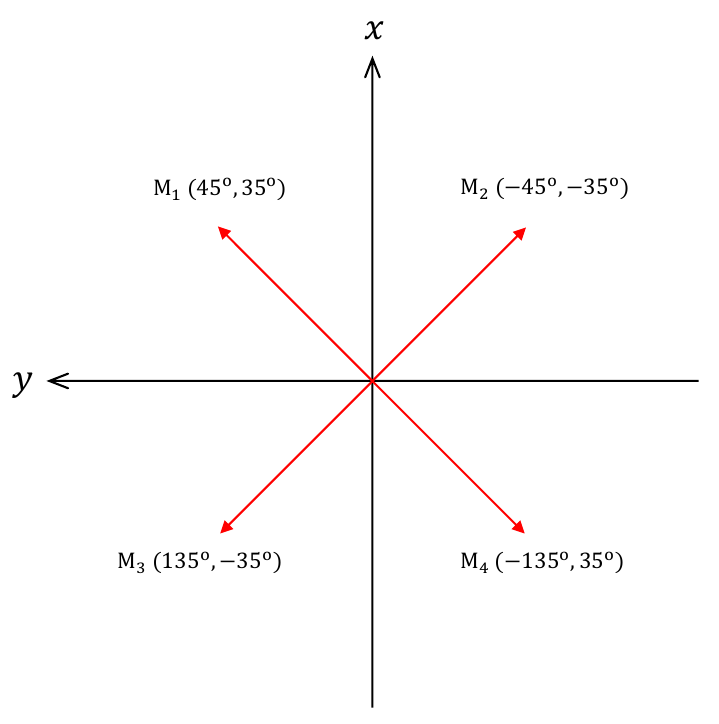}}
  \caption{Top view of an arrangement of the four microphones in spherical coordinates from the z-axis for the MIC format data set. Note that the azimuth angle is increasing counter-clockwise. $\rm M_m$($\phi_m$, $\theta_m$) denotes the azimuth and elevation angles for the $m$-{\upshape th} microphone.}
  \label{fig:mic}
\end{figure}

The development data set for the SELD task of the DCASE 2020 Challenge \cite{politis2020dataset} contains multiple spatial sound-scene recordings generated by convolving randomly chosen isolated sound event samples with real-life room impulse responses (RIRs) collected by a spherical microphone array called EM32 Eigenmike \cite{acoustics2013em32}. Each scene recording is delivered in two 4-channel spatial recording formats, MIC and FOA. The MIC data set is extracted directly by selecting four channels of the spherical microphone array, corresponding to a tetrahedral capsule arrangement, as shown in Fig.~\ref{fig:mic}. For these four microphones mounted on a spherical baffle, an analytical expression for the directional array response is given by the expansion for the MIC format \cite{politis2020dataset}:
\begin{equation}
\label{h_mic}
\begin{split}
& H_m^{\rm{MIC}}(\phi_m, \theta_m, \phi, \theta, \omega)= \\
& \frac{1}{(\omega R/c)^2}\sum_{n=0}^{30}\frac{j^{n-1}}{h_n^{'(2)}(\omega R/c)}(2n+1)P_n({\rm cos}(\gamma_m)) ,
\end{split}
\end{equation}
where the directional response for each channel is expanded to 30 terms according to \cite{politis2020dataset}. $R=4.2\,$cm is the spherical microphone array radius. The channel index is denoted by $m$. Symbols $\phi$ and $\theta$ denote the azimuth and elevation angles of the sound source, whereas $\phi_m$ and $\theta_m$ denote the azimuth and elevation angles of $m$-th microphone as shown in Fig.~\ref{fig:mic}. The angular frequency is denoted by $\omega$ and can be expressed as $\omega=2\pi f$. The microphone array radius and sound speed are denoted by $R$ and $c$, respectively. $j$ denotes the imaginary unit. $h_n^{'(2)}$ is the derivative with respect to the parameter passed to the second kind of spherical Hankel function. $P_n$ denotes the $n$-th order non-normalized Legendre polynomial. And $\gamma_m$ denotes the angle between the $m$-th microphone position and the sound source. From Eq.~(\ref{h_mic}), we can see that the directional response of the $m$-th channel is a function of the cosine angle between the microphone position and the DOA:
\begin{equation}
\label{cos}
{\rm cos}(\gamma_m) = {\rm sin}(\theta){\rm sin}(\theta_m){\rm cos}(\phi-\phi_m)+{\rm cos}(\theta){\rm cos}(\theta_m) .
\end{equation}

Ambisonics is another data format that decomposes a sound field on an orthogonal basis of spherical harmonic functions. In this study, first-order decomposition is used to generate the FOA data set. It is obtained by converting the spherical microphone array signals to first-order Ambisonics with measurement-based encoding filters \cite{politis2017comparing}. The FOA signal consists of four channels ($W,Y,Z,X$) with $W$ corresponding to an omnidirectional microphone and ($Y,Z,X$) corresponding to three bidirectional microphones aligned on the Cartesian axes. With $t$ and $f$ as the T-F bin indices, considering there is a point source $p(t,f)$ from DOA in the short-time Fourier transform (STFT) domain given by azimuth angle $\phi$ and elevation angle $\theta$, the sound field on the four FOA channels can be decomposed as
\begin{equation}
\left[ \begin{array}{c} W(t,f) \\ Y(t,f) \\ Z(t,f) \\ X(t,f) \\ \end{array} \right]=\left[ \begin{array}{c} 1 \\ \rm{sin}(\phi) \rm{cos}(\theta) \\ \rm{sin}(\theta) \\ \rm{cos}(\phi) \rm{cos}(\theta) \\ \end{array} \right]p(t,f) .
\end{equation}
Using the SN3D normalization scheme of Ambisonices \cite{daniel2000representation}, the frequency-independent directional response of the $m$-th channel $H_m^{\rm{FOA}}$($\phi,\theta,f$) for FOA is given by
\begin{align}
\label{h_foa1}
H_1^{\rm{FOA}}(\phi,\theta,f)&=1 ,\\
H_2^{\rm{FOA}}(\phi,\theta,f)&=\rm{sin}(\phi) \rm{cos}(\theta) ,\\
H_3^{\rm{FOA}}(\phi,\theta,f)&=\rm{sin}(\theta) ,\\
H_4^{\rm{FOA}}(\phi,\theta,f)&=\rm{cos}(\phi) \rm{cos}(\theta) .
\label{h_foa2}
\end{align}

New DOA representations can be generated based on the spatial responses of the MIC and FOA data sets by applying transformations to audio channels. For data with the MIC format, not only level but also time differences are encoded, thus the spatial responses of the augmented data must be the same as that of the original data. Only level differences are encoded for the FOA format data, which means that there may exist sign inversion for the spatial responses of the augmented data. There are only a limited set of transformations that can be applied to the audio channels in order to keep the spatial responses of the MIC data unchanged. Specifically, channel swapping is used for the MIC data and there are only eight allowable transformations to obtain effective audio data and the corresponding DOA representations. To obtain the same DOA labels for the FOA data, channel transformations can be applied to the FOA channels according to the spatial responses.

\begin{table*}[t]
\caption{{The ACS augmentation approach for both MIC and FOA data sets. $C_m$ and $C_m^{\rm new}$ denote the $m$-{\upshape th} channel data of the original and augmented data sets, respectively.}}
\label{ACS}
\centering
\begin{tabular}{|c|c|c|}
\hline
\rule{0pt}{9pt} DOA Transformation &MIC Dataset &FOA Dataset \\
\hline
\rule{0pt}{10pt} \makecell*[c]{$\phi=\phi-\pi/2,\theta=-\theta$}  &$C_1^{\rm new}=C_2,C_2^{\rm new}=C_4,C_3^{\rm new}=C_1,C_4^{\rm new}=C_3$   &$C_1^{\rm new}=C_1,C_2^{\rm new}=-C_4,C_3^{\rm new}=-C_3,C_4^{\rm new}=C_2$   \\
\hline
\rule{0pt}{10pt} \makecell*[c]{$\phi=-\phi-\pi/2,\theta=\theta$}  &$C_1^{\rm new}=C_4,C_2^{\rm new}=C_2,C_3^{\rm new}=C_3,C_4^{\rm new}=C_1$   &$C_1^{\rm new}=C_1,C_2^{\rm new}=-C_4,C_3^{\rm new}=C_3,C_4^{\rm new}=-C_2$   \\
\hline
\rule{0pt}{10pt} \makecell*[c]{$\phi=\phi,\theta=\theta$}         &$C_1^{\rm new}=C_1,C_2^{\rm new}=C_2,C_3^{\rm new}=C_3,C_4^{\rm new}=C_4$  &$C_1^{\rm new}=C_1,C_2^{\rm new}=C_2,C_3^{\rm new}=C_3,C_4^{\rm new}=C_4$   \\
\hline
\rule{0pt}{10pt} \makecell*[c]{$\phi=-\phi,\theta=-\theta$}       &$C_1^{\rm new}=C_2,C_2^{\rm new}=C_1,C_3^{\rm new}=C_4,C_4^{\rm new}=C_3$   &$C_1^{\rm new}=C_1,C_2^{\rm new}=-C_2,C_3^{\rm new}=-C_3,C_4^{\rm new}=C_4$   \\
\hline
\rule{0pt}{10pt} \makecell*[c]{$\phi=\phi+\pi/2,\theta=-\theta$}  &$C_1^{\rm new}=C_3,C_2^{\rm new}=C_1,C_3^{\rm new}=C_4,C_4^{\rm new}=C_2$   &$C_1^{\rm new}=C_1,C_2^{\rm new}=C_4,C_3^{\rm new}=-C_3,C_4^{\rm new}=-C_2$   \\
\hline
\rule{0pt}{10pt} \makecell*[c]{$\phi=-\phi+\pi/2,\theta=\theta$}  &$C_1^{\rm new}=C_1,C_2^{\rm new}=C_3,C_3^{\rm new}=C_2,C_4^{\rm new}=C_4$   &$C_1^{\rm new}=C_1,C_2^{\rm new}=C_4,C_3^{\rm new}=C_3,C_4^{\rm new}=C_2$   \\
\hline
\rule{0pt}{10pt} \makecell*[c]{$\phi=\phi+\pi,\theta=\theta$}    &$C_1^{\rm new}=C_4,C_2^{\rm new}=C_3,C_3^{\rm new}=C_2,C_4^{\rm new}=C_1$   &$C_1^{\rm new}=C_1,C_2^{\rm new}=-C_2,C_3^{\rm new}=C_3,C_4^{\rm new}=-C_4$   \\
\hline
\rule{0pt}{10pt} \makecell*[c]{$\phi=-\phi+\pi,\theta=-\theta$}    &$C_1^{\rm new}=C_3,C_2^{\rm new}=C_4,C_3^{\rm new}=C_1,C_4^{\rm new}=C_2$   &$C_1^{\rm new}=C_1,C_2^{\rm new}=C_2,C_3^{\rm new}=-C_3,C_4^{\rm new}=-C_4$   \\
\hline
\end{tabular}
\end{table*}

\begin{figure}
  \centering
  \centerline{\includegraphics[width=0.9\columnwidth]{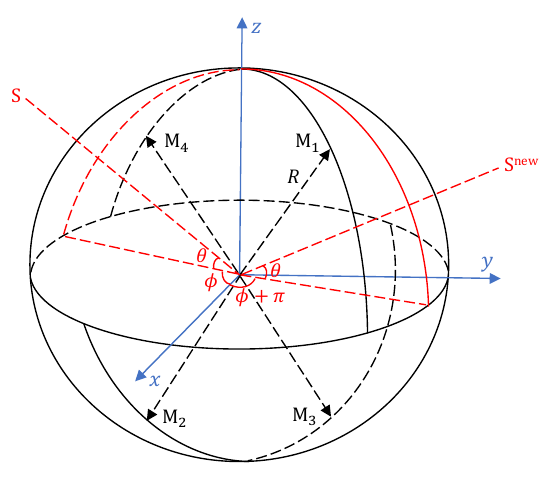}}
  \caption{One DOA transformation example: ($\phi=\phi+\pi, \theta=\theta$). $\rm {M_1, M_2, M_3}$, and $\rm {M_4}$ are four microphones in MIC dataset. $\rm R$ is the spherical microphone array radius. S denotes the original DOA with azimuth angle $\phi$ and elevation angle $\theta$, and $\rm {S^{new}}$ denotes the new source with angles $\phi+\pi$ and $\theta$.}
  \label{fig:transf}
\end{figure}

Table~\ref{ACS} lists all eight DOA transformations (including identity) for ACS spatial augmentation. Take one transformation, $\phi=\phi+\pi, \theta=\theta$, as shown in Fig.~\ref{fig:transf} for example. $\rm {M_1, M_2, M_3}$, and $\rm {M_4}$ are four microphones arranged on a spherical baffle to extract the MIC data. The azimuth and elevation angles of the four microphones are shown in Fig.~\ref{fig:mic}. Considering there is an original sound source $\rm S$ from DOA given by azimuth angle $\phi$ and elevation angle $\theta$, then the MIC format data can be denoted as ($\rm {C_1, C_2, C_3, C_4}$), which means that the $m$-th channel data $\rm{C}_m$ is extracted by the $m$-th microphone $\rm{M}_m$. By applying a DOA transformation, the newly generated sound source $\rm {S^{new}}$ has an azimuth angle $\phi+\pi$ and an elevation angle $\theta$. It can be seen in Fig.~\ref{fig:transf} that the relative location relationship, between the newly generated sound source $\rm {S^{new}}$ and spherical microphone array, stays unchanged. Due to the symmetry of the four-microphone arrangement, it is equivalent to obtain multi-channel data ($\rm {C_4, C_3, C_2, C_1}$) for sound source $\rm {S^{new}}$, corresponding to swapping the first and fourth channels plus the second and third channels. From a theoretical perspective, after applying the DOA transformation ($\phi=\phi+\pi, \theta=\theta$), the spatial response of each channel for both the MIC and FOA data can be calculated according to Eqs. (\ref{cos}) and (\ref{h_foa1}-\ref{h_foa2}) to generate the augmented data accordingly. Note that ACS for the FOA data is also discussed in \cite{MazzonYasuda2019}.

The ACS augmentation approach is simple to implement. It can be applied to any sound event sample, whether non-overlapping or overlapping, whether static or moving, by directly performing transformations on audio channels. The original DOA labels are limited in the domain that azimuth $\phi \in [-180^{\rm o}, 180^{\rm o}]$ and elevation $\theta \in [-90^{\rm o}, 90^{\rm o}]$. After applying DOA transformations, it is easy to control the augmented DOA labels in the same domain.

\subsection{Multi-Channel Simulation (MCS)}
\label{subsec:MCS}

The MCS spatial augmentation method is proposed to deal with small SELD datasets based on a beamforming approach. Unlike ACS, the MCS method does not rely on the characteristics of microphone arrays, so it is more widely applicable.

\begin{figure}
  \centering

  \begin{minipage}[t]{0.453\linewidth}
  \centering
  \includegraphics[width=1\linewidth]{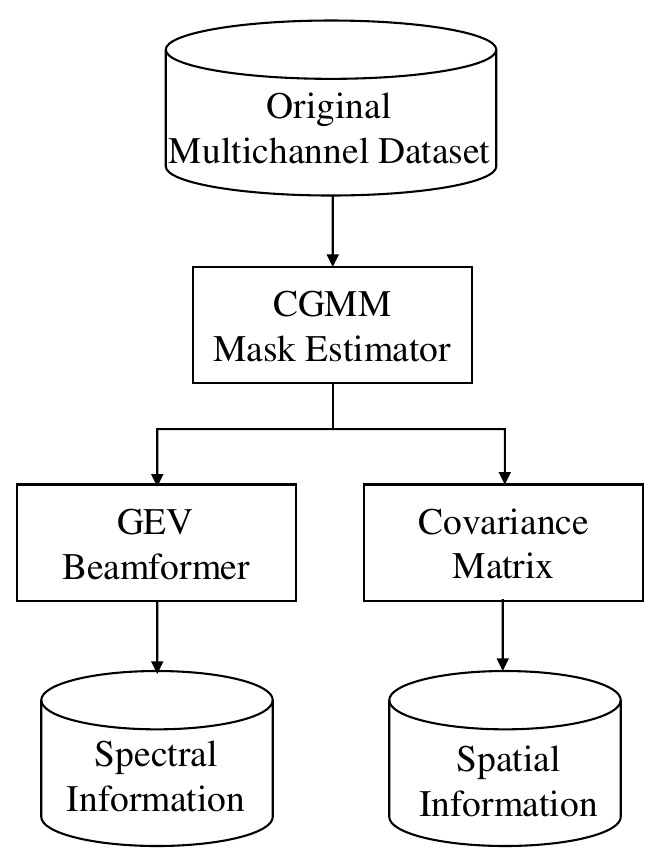}
  \centerline{\footnotesize{(a) Step 1}}
  \end{minipage}
  \label{fig:mds1}
  \qquad
  \begin{minipage}[t]{0.453\linewidth}
  \centering
  \includegraphics[width=1\linewidth]{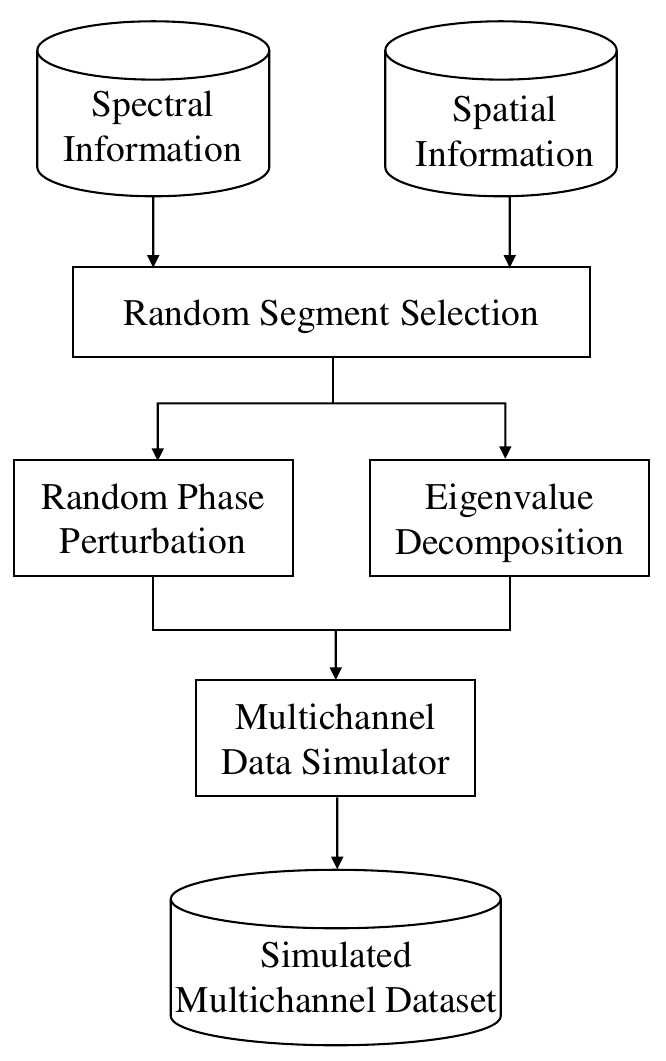}
  \centerline{\footnotesize{(b) Step 2}}
  \end{minipage}
  \label{fig:mds2}

  \centering
  \caption{Proposed multi-channel simulation (MCS) work flow.}
  \label{fig:MCS}
\end{figure}

Sound-scene recordings are delivered in multi-channel data containing both spectral and spatial information. Spectral and spatial information are related to the category and DOA of sound event, respectively. We propose a novel MCS augmentation technique with the motivation of increasing the number of DOA representations for audio segments containing non-overlapping and non-moving sound events. And this study is focused on augmentation from original data, without using external data. MCS consists of two steps as shown in Fig.~\ref{fig:MCS}. The first step involves extracting spectral and spatial information for all audio segments containing isolated and static sound events. We use the CGMM model, a mask-based beamforming approach originally proposed for ASR in \cite{higuchi2016robust}, to estimate T-F masks that represent the probabilities of the T-F units belonging to a sound source or only noise. In \cite{warsitz2007blind}, the authors demonstrated that the GEV beamformer applied with a single-channel post-filter could generate distortionless speech. Here we adopt such a GEV beamformer to extract the desired spectral vector. It should be noted that the spatial vector is estimated by calculating the covariance matrix of the sound source. Two sets containing spectral and spatial information respectively are generated as shown in the first step of Fig.~\ref{fig:MCS}.

In the second step, a pair of spectral and spatial information is randomly selected from these two sets generated in the first step to simulate new multi-channel data. Eigenvalue decomposition of the spatial vector is performed to calculate the eigenvalue and eigenvector. And a random phase perturbation is performed on the spectral vector to guarantee a full-rank covariance matrix. Finally, a multi-channel simulator is adopted to generate simulated multi-channel data by combining the eigenvalue, eigenvector and spectral vector with random phase.

In order to exploit the spatial features of both MIC and FOA data, we concatenate the two 4-channel spatial formats, resulting in an 8-channel signal. First, we collect all audio segments of signals in the original dataset containing non-overlapping and non-moving sound events according to SED and DOA labels. Considering $s(l)$ is the $l$-th sample of an audio segment containing static sound event in time domain, we have an array of $M$=8 microphones in total to record sound samples, thus the $l$-th sample of the observed audio segment at the $m$-th microphone can be written as
\begin{equation}
x_m(l)=\sum_{\tau}h_m(\tau)s(l-\tau)+n_m(l) ,
\end{equation}
where $n_m(l)$ denotes the $l$-th sample of the audio segment of the noise signal recorded at $m$-th microphone, and $h_m(\tau)$ denotes an impulse response between the sound source and the $m$-th microphone. Via STFT, the microphone array observation vector transformed into T-F domain is given by

\begin{equation}
{\rm\mathbf x}_{f,t}={\rm\mathbf h}_f S_{f,t}+{\rm\mathbf n}_{f,t} ,
\end{equation}

with
\begin{align}
{\rm\mathbf x}_{f,t}&=[X_{f,t,1}, X_{f,t,2}, ..., X_{f,t,M}]^{\rm T} ,\\
{\rm\mathbf h}_f    &=[H_{f,1}, H_{f,2}, ..., H_{f,M}]^{\rm T} ,\\
{\rm\mathbf n}_{f,t}&=[N_{f,t,1}, N_{f,t,2}, ..., N_{f,t,M}]^{\rm T} ,
\end{align}
where ${\rm\mathbf x}_{f,t}$, ${\rm\mathbf h}_f$, and ${\rm\mathbf n}_{f,t}$ are mixture vector, steering vector, and noise vector, respectively, with $f$ denoting frequency index and $t$ denoting frame index. $S_{f,t}$ is the spectrogram of target sound source and $[\cdot]^{\rm T}$ denotes non-conjugate transposition.

We use a CGMM-based method proposed in \cite{higuchi2016robust} to estimate T-F masks representing the probabilities of the T-F units belonging to a sound source or only noise. Using CGMM, the observed mixture vector can be clustered into either one containing a sound source or the other containing only noise, and expressed as
\begin{equation}
{\rm\mathbf x}_{f,t}={\rm\mathbf h}_{f}^{(v)}S_{f,t}^{(v)} \,\,\,({\rm where}\,d_{f,t}=v) ,
\end{equation}
where $d_{f,t}$ denotes the category index at the time frame $t$ and frequency bin $f$. When $v=s$, the category represents a sound source. When $v=n$, the category represents noise. $S_{f,t}^{(v)}\sim \mathcal{N}_c(0,\sigma_{f,t}^{(v)})$ is assumed to follow a complex Gaussian distribution, then the observed multi-channel vector is assumed to follow a multivariate complex Gaussian distribution
\begin{equation}
{\rm\mathbf x}_{f,t}|d_{f,t}=v \sim \mathcal{N}_c({\mathbf 0},\sigma_{f,t}^{(v)}{\rm\mathbf H}_{f}^{(v)}) ,
\end{equation}
where
\begin{align}
{\rm\mathbf H}_{f}^{(v)}&={\rm\mathbf h}_{f}^{(v)}({\rm\mathbf h}_{f}^{(v)})^{\rm H} ,\\
\mathcal{N}_c({\rm\mathbf x}|\boldsymbol{\mu},{\mathbf \Sigma})&=\frac{1}{|\pi{\mathbf \Sigma}|}\exp\left(-({\rm\mathbf x}-{\boldsymbol \mu})^{\rm H}{\mathbf \Sigma}^{-1}({\rm\mathbf x}-{\boldsymbol \mu})\right) ,
\end{align}
with $(\cdot)^{\rm H}$ denoting conjugate transposition. $\sigma_{f,t}^{(v)}$ and ${\rm\mathbf H}_{f}^{(v)}$ are two CGMM parameters that can be estimated using a maximum likelihood (ML) criterion. Through expectation-maximization (EM), the T-F masks can be updated as follows:
\begin{equation}
\lambda_{f,t}^{(v)}\leftarrow \frac{p({\rm\mathbf x}_{f,t}|d_{f,t}=v,{\mathbf 0}, \sigma_{f,t}^{(v)}{\rm\mathbf H}_{f}^{(v)})}{\sum_vp({\rm\mathbf x}_{f,t}|d_{f,t}=v,{\mathbf 0}, \sigma_{f,t}^{(v)}{\rm\mathbf H}_{f}^{(v)})} .
\end{equation}
The probability of T-F unit $(f,t)$ belonging to a sound source or only noise can be measured by $\lambda_{f,t}^{(v)}$ after convergence.

The spatial information of a static sound source is contained in the multi-channel audio segment. We estimate it by calculating the covariance matrix of the enhanced sound source, written as:
\begin{equation}
\mathcal{R}_f^{(s)}=\frac{1}{\sum_t\lambda_{f,t}^{(s)}}\sum_t\lambda_{f,t}^{(s)}{\rm\mathbf x}_{f,t}{\rm\mathbf x}_{f,t}^{\rm H} ,
\end{equation}
where $\lambda_{f,t}^{(s)}$ denotes the probability of the T-F unit $(t,f)$ belonging to a sound source. Finally, we perform an energy normalization on the covariance matrix $\mathcal{R}_f^{(s)}$ to extract spatial vector $\mathcal{S}_f$ as follows
\begin{equation}
\mathcal{S}_f=\frac{M\mathcal{R}_f^{(s)}}{{\rm tr}(\mathcal{R}_f^{(s)})} .
\end{equation}

To estimate the spectral vector, we adopt a GEV beamformer and a single-channel post-filter \cite{warsitz2007blind}. Our goal is to find a vector of optimal filter coefficients ${\rm\mathbf w}_f=[W_{f,1}, W_{f,2}, ..., W_{f,M}]^{\rm T}$ with which the beamformer output achieves the maximum signal-to-noise (SNR) ratio and is distortionless at the same time. The output can be written as
\begin{equation}
{\hat S}_{f,t}={\rm\mathbf w}_{f}^{\rm H}{\rm\mathbf x}_{f,t} .
\end{equation}

According to \cite{warsitz2007blind}, the filter coefficients of the GEV beamformer ${\rm\mathbf w}_{{\rm SNR},f}$ is the eigenvector corresponding to the largest eigenvalue of $(\Phi_f^{\rm\mathbf n})^{-1}\Phi_f^{\rm\mathbf x}$, where $\Phi_f^{\rm\mathbf x}$ and $\Phi_f^{\rm\mathbf n}$ denote the cross power spectral density (PSD) matrices of the observed mixture and the noise, respectively.

The optimal coefficients vector of the GEV beamformer is computed by maximizing the output SNR, which may introduce speech distortion. To obtain a distortionless target sound source, a single-channel post-filter $\omega_f$ is added as follows
\begin{equation}
{\rm\mathbf w}_f=\omega_f{\rm\mathbf w}_{{\rm SNR},f} .
\end{equation}
According to the blind analytical normalization method \cite{warsitz2007blind}, the post-filter $\omega_f$ is obtained as
\begin{equation}
\omega_f=\frac{\sqrt{{\rm\mathbf w}_{{\rm SNR},f}^{\rm H}\Phi_f^{\rm\mathbf n}\Phi_f^{\rm\mathbf n}{\rm\mathbf w}_{{\rm SNR},f}/M}}{{\rm\mathbf w}_{{\rm SNR},f}^{\rm H}\Phi_f^{\rm\mathbf n}{\rm\mathbf w}_{{\rm SNR},f}} .
\end{equation}

So far we generate the spectral and spatial information collected in two sets as shown in the first step of Fig.~\ref{fig:MCS} for all audio segments containing non-overlapping and non-moving sound events. In the second step, spectral information ${\hat S}_{f,t}$ and spatial information $\mathcal{S}_f$ are selected from these two sets randomly shuffled to simulate multi-channel audio segment. For the spatial case, it is obvious that $\mathcal{S}_f=\mathcal{S}_f^{\rm H}$. The eigenvalue decomposition of such a conjugate matrix $\mathcal{S}_f$ can be written as follows
\begin{equation}
\label{spatial}
\mathcal{S}_f={\rm\mathbf U}_f {\mathbf \Lambda}_f {\rm\mathbf U}_f^{\rm H} ,
\end{equation}
where ${\mathbf \Lambda}_f={\rm diag}(\lambda_{f,1},\lambda_{f,2},...,\lambda_{f,M})$ denote the eigenvalues, and ${\rm\mathbf U}_f=[{\rm\mathbf u}_{f,1},{\rm\mathbf u}_{f,2},...,{\rm\mathbf u}_{f,M}]^{\rm T}$ are the corresponding eigenvectors. Eq.~(\ref{spatial}) can also be expressed as a sum of $M$ components:
\begin{equation}
\mathcal{S}_f=\sum_{m=1}^M \lambda_{f,m}{\rm\mathbf u}_{f,m}{\rm\mathbf u}_{f,m}^{\rm H} .
\end{equation}

Now we have spectral information ${\hat S}_{f,t}$ extracted from one audio segment and spatial information $\mathcal{S}_f$ extracted from another. Our goal is to simulate a new audio segment. Its SED label corresponds to that of the audio segment extracting ${\hat S}_{f,t}$ while its DOA label corresponds to that of the audio segment extracting $\mathcal{S}_f$. The simulated multi-channel audio segment can be written as
\begin{equation}
\label{simu}
{\hat {\rm\mathbf x}}_{f,t}=\sum_{m=1}^M\sqrt{\lambda_{f,m}}{\hat S}_{f,t}{\rm exp}(2\pi \xi_{t,m} j){\rm\mathbf u}_{f,m} ,
\end{equation}
where $\xi_{t,1}=0,\xi_{t,m}\sim U(0,1),m=2,3,...,M$, and $U(0,1)$ denotes a standard uniform distribution. $j$ denotes the imaginary unit. As $\xi_{t,m}$ is randomly chosen for all the channels except for the first one, the term ${\rm exp}(2\pi \xi_{t,m} j)$ could be regarded as a random phase perturbation factor with respect to the enhanced sound source ${\hat S}_{f,t}$ in the first channel, aiming to simulate diffusing characteristics of reverberation and noise in real spatial sound-scene. It can make the covariance matrix of the simulated audio segment full-rank, which has been demonstrated as a more appropriate model in realistic reverberant environments \cite{duong2010under}. The random phase perturbation factor does affect the phase differences between channels, however, it is simple to prove that the covariance matrix of $\hat{\rm\mathbf x}_{f,t}$ is independent with phase perturbation. This shows that the simulated segment contains expected spatial information, which demonstrates that random phase perturbation does not affect the observed locations of sound events. Multi-channel audio segment in the time domain is simulated by applying inverse STFT to Eq.~(\ref{simu}). As we use audio clips with a fixed length to train models, the simulated segments are spliced into a long sample with the length of 60 seconds.

\subsection{Time-Domain Mixing (TDM)}
\label{subsec:TDM}

When two sound events occur close to each other in time, it is more difficult to perform SED and SSL. To improve the generalization of our model to handle overlapping sources, we perform TDM for two non-overlapping sources. The SELD labels for the augmented data are the union of the original labels for the two sound events. Although no new DOA is generated, TDM increases the number of overlapping training samples, which has proved to be effective.

\subsection{Time-Frequency Masking (TFM)}
\label{subsec:TFM}

SpecAugment is a simple yet helpful augmentation method in ASR \cite{park2019specaugment}. We find it useful for SED but it may sometimes cause performance degradation for SSL. Nevertheless we found SpecAugment to be effective for both SED and SSL with a large-sized training set. In this study, masks are applied to the time and frequency dimensions randomly for each input log Mel-spectrogram feature in each batch during training.

\section{ResNet-Conformer Based SELD System}
\label{sec:conformer}

In our submitted system for the DCASE 2020 Challenge \cite{Du2020_task3_report,wang2021model}, we investigated several deep learning based acoustic models for the SELD task, which consist of high-level feature representation, temporal context representation and full connection. The high-level feature representation module contains a series of CNN blocks, each having a 2D convolution layer followed by a batch normalization process, a rectified linear unit (Relu), and a max-pooling operation. The temporal context representation module models the temporal structures within sound events. We use two parallel branches in the fully-connected (FC) module to perform SED and SSL simultaneously, similar to the official baseline SELD system \cite{politis2020dataset}. Moreover, we used modified versions of ResNet \cite{he2016deep} and Xception \cite{chollet2017xception} to learn local shift-invariant features. Besides the bidirectional gated recurrent unit (GRU) used in the baseline system, we also adopted factorized time delay neural network (TDNN-F) \cite{povey2018semi} to exploit longer temporal context dependency in the audio signal.

\begin{figure}
  \centering
  \centerline{\includegraphics[width=0.88\columnwidth]{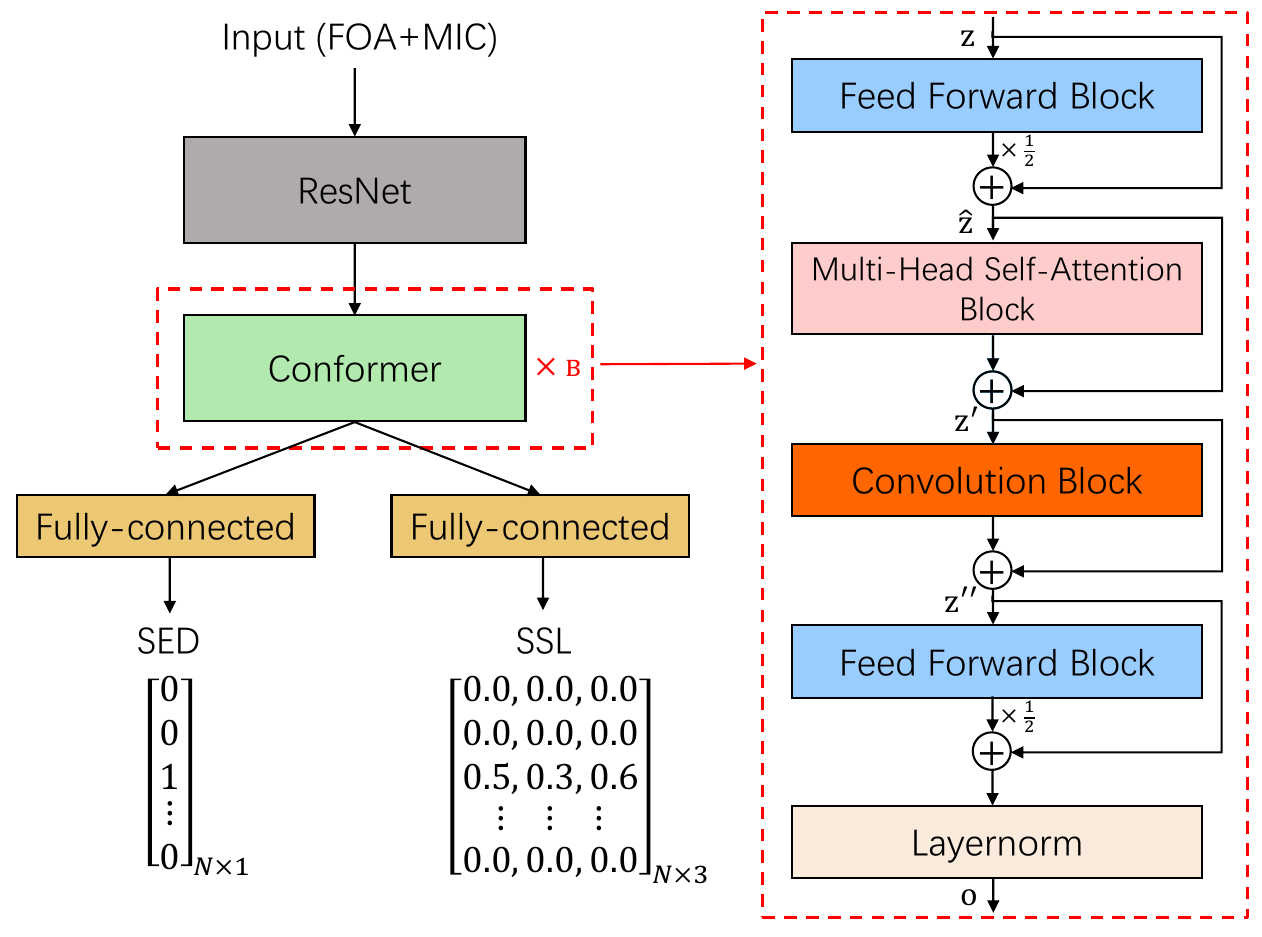}}
  \caption{A flow chart of the proposed ResNet-Conformer architecture for the SELD task and a detailed implementation of the Conformer module.}
  \label{fig:conformer}
\end{figure}

The Conformer architecture, combining convolution and transformer layers \cite{vaswani2017attention}, was proposed in \cite{gulati2020conformer} and achieved state-of-the-art results for ASR. Its applications were then extended to continuous speech separation \cite{chen2020continuous}, sound event detection and separation in domestic environments \cite{Miyazaki2020}. The convolution layers are effective to extract local fine-grained features while the transformer models are good at capturing long-range global context. Thus Conformer is supposed to be able to model both local and global context dependencies in an audio sequence. In this paper, we examine the use of Conformer for the SELD task. We use ResNet to extract local shift-invariant features. Then Conformer is adopted to learn both local and global context representations. We call our acoustic model ResNet-Conformer. Fig.~\ref{fig:conformer} shows an overview of the proposed architecture for the SELD task and a detailed Conformer implementation.  As shown in the left panel in Fig.~\ref{fig:conformer}, two parallel branches contain two FC layers, each performing individual SED and SSL subtasks. Note that $N$ is equal to the number of sound event classes.

Shown in the right dashed box in Fig.~\ref{fig:conformer}, Conformer is composed of two feed forward blocks that sandwich a multi-head self-attention (MHSA) block and a convolution block. Relative positional encoding scheme is used in the Conformer architecture. All these four blocks start with layer normalization operations. And the second feed forward block is also followed by layer normalization. A residual connection is added behind each block. Assume ${\mathbf z}$ is the input to Conformer, the output ${\mathbf o}$ can be calculated through intermediate ${\hat{\mathbf z}}$, ${\mathbf z}'$ and ${\mathbf z}''$ as
 \begin{align}
 {\hat{\mathbf z}}&={\mathbf z}+\frac{1}{2}{\rm FFN}(\mathbf z)  ,\\
 {\mathbf z}'&={\hat{\mathbf z}}+{\rm MHSA}({\hat{\mathbf z}})  ,\\
 {\mathbf z}''&={\mathbf z}'+{\rm Conv}({\mathbf z}')  ,\\
 {\mathbf o}&={\rm Layernorm}({\mathbf z}''+\frac{1}{2}{\rm FNN}({\mathbf z}'')) ,
 \end{align}
where ${\rm FFN}(\cdot)$, ${\rm MHSA}(\cdot)$, ${\rm Conv}(\cdot)$ and ${\rm Layernorm}(\cdot)$ denote a feed forward network block, a multi-head self-attention block, a convolution block, and a layer normalization process, respectively. The input $\hat{\mathbf z}$ to MHSA is first processed by layer normalization and then converted to the query ${\rm\mathbf Q}$, key ${\rm\mathbf K}$, and value ${\rm\mathbf V}$ by performing linear projection as follows:
\begin{equation}
{\rm Multihead}({\rm\mathbf Q},{\rm\mathbf K},{\rm\mathbf V})={\rm Concat}({\rm\mathbf H}_1,{\rm\mathbf H}_2,...,{\rm\mathbf H}_h){\rm\mathbf W}^O ,
\end{equation}
\begin{equation}
{\rm\mathbf H}_i={\rm Attention}({\rm\mathbf Q_i},{\rm\mathbf K_i},{\rm\mathbf V_i}) ,
\end{equation}
\begin{equation}
{\rm Attention}({\rm\mathbf Q_i},{\rm\mathbf K_i},{\rm\mathbf V_i})={\rm Softmax}\left(\frac{{\rm\mathbf Q_i}{\rm\mathbf K_i}^{\rm T}}{\sqrt{d_k}}\right){\rm\mathbf V_i} ,
\end{equation}
where $\hat{\mathbf z}' = {\rm Layernorm}({\hat{\mathbf z}}), {\rm\mathbf Q}=\hat{\mathbf z}'{\rm\mathbf W}^{Q}, {\rm\mathbf K}=\hat{\mathbf z}'{\rm\mathbf W}^{K}, {\rm\mathbf V}=\hat{\mathbf z}'{\rm\mathbf W}^{V}$, $h$ denotes the number of the attention heads. ${\rm\mathbf W}^{Q}, {\rm\mathbf W}^{K} \in \mathbb{R}^{d\times (d_k \times h)}, {\rm\mathbf W}^{V} \in \mathbb{R}^{d\times (d_v \times h)}$ are learnable parameter matrices. Variables ${\rm\mathbf Q_i},{\rm\mathbf K_i}$, and ${\rm\mathbf V_i}$ denote the query, key, and value of self-attention for the $i$-th head, respectively. ${\rm\mathbf W}^{O} \in \mathbb{R}^{(d_v \times h) \times d}$ is the final linear parameter matrix applied on the concatenated feature vector. $d$ denotes the dimension of input vector. $d_k$ and $d_v$ denote the dimensions of key and value for each head, respectively. Concat$(\cdot)$ denotes the concatenation operation.

\begin{figure}
  \centering
  \centerline{\includegraphics[width=0.75\linewidth]{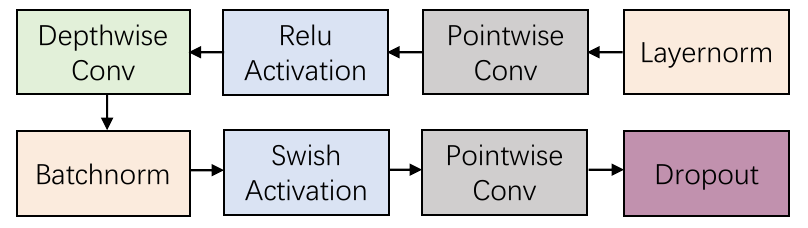}}
  \caption{A detailed implementation of the convolution block.}
  \label{fig:conv}
\end{figure}

\begin{figure}
  \centering
  \centerline{\includegraphics[width=0.98\linewidth]{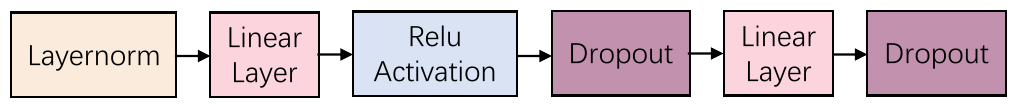}}
  \caption{A detailed implementation of the feed forward network block.}
  \label{fig:ffn}
\end{figure}

${\rm Conv}(\cdot)$ is illustrated in Fig.~\ref{fig:conv} which contains two pointwise convolution layers sandwiching a depthwise convolution layer. The first pointwise convolution layer is followed by Relu activation and the second convolution layer is followed by a dropout operation. Following the depthwise convolution layer is a batch normalization and a Swish activation. The feed forward network consists of two linear layers and a nonlinear activation Relu in between as illustrated in Fig.~\ref{fig:ffn}. Dropout is used to help regularize the network.

Multitask learning is used to train the ResNet-Conformer model as shown in Fig.~\ref{fig:conformer}. The output layers in the two branches consist of multiple targets to be predicted, including active sound classes and the corresponding DOAs. Sigmoid activation function is used for the SED branch and tanh activation function is used for the SSL branch. A joint loss function is adopted to solve the SED and SSL subtasks simultaneously. The SED subtask is performed as a multi-label classification with a binary cross-entropy (BCE) loss. The SSL subtask is performed as a multi-output regression with a masked mean squared error (MSE) loss \cite{cao2019polyphonic}. The masked MSE loss is computed based on the ground truth of each sound event class, hence not contributing to the training when a sound event is not active. The multi-objective loss function to be minimized can be expressed as
\begin{equation}
\begin{split}
L=&-\frac{\alpha_1}{T}\sum_t\sum_ny_{t,n}^{\rm SED}{\rm log}{\hat {y}}_{t,n}^{\rm {SED}} \\
&+\frac{\alpha_2}{T}\sum_t\sum_n||(\hat {\mathbf y}_{t,n}^{\rm {SSL}}-{\mathbf y}_{t,n}^{\rm SSL})y_{t,n}^{\rm SED}||^2 ,
\end{split}
\end{equation}
where $\hat {y}_{t,n}^{\rm SED}$ and $\hat {\mathbf y}_{t,n}^{\rm SSL}$ are the active probability estimation and DOA estimation for the $n$-th sound event at the $t$-th frame, respectively. Correspondingly, $y_{t,n}^{\rm SED}$ and $\mathbf {y}_{t,n}^{\rm SSL}$ are the reference versions. Both $\hat {\mathbf y}_{t,n}^{\rm SSL}$ and $\mathbf {y}_{t,n}^{\rm SSL}$ are 3-dimensional Cartesian representations of the DOA. $T$ denotes the total frame number in a minibatch. The SED classification loss and SSL regression loss are combined for joint optimization during training with loss weights $\alpha_1$ equal to 1 and $\alpha_2$ equal to 10.

\section{Experimental Results and Discussions}
\label{sec:exp}

We evaluate SELD on the official development sets of Task 3 in the DCASE 2020 and 2022 Challenges \cite{politis2020dataset,politis2022starss22}. The data set of SELD task in DCASE 2020 Challenge contains 600 60-second audio recordings with a 24 kHz sampling rate. They are divided into six equal-sized splits with four splits for training, one split for validation and the last split for testing. The model that performs best on the validation set is selected for testing. There are 14 sound classes of spatial events as listed in Table~\ref{class}. The four proposed data augmentation approaches as described in Section~\ref{sec:DA} are used to expand the training data set.

The data set of SELD task in DCASE 2022 Challenge contains 121 audio recordings with the duration from 30 seconds to 5 minutes. The main difference between these two challenge tasks is that dataset in DCASE 2020 Challenge is computationally generated while dataset in DCASE 2022 Challenge is collected in real sound scenes. All ablation experiments shown in Section~\ref{sec:exp}-A to Section~\ref{sec:exp}-D are conducted on the SELD data set of DCASE 2020 Challenge. Section~\ref{sec:exp}-E lists experimental results on the data set of DCASE 2022 Challenge.
\begin{table*}[htb]\scriptsize
\caption{The 14 sound classes of the spatial events in the DCASE 2020 Challenge.}
\label{class}
\centering
\begin{tabular}{|m{0.68cm}<{\centering}|m{0.68cm}<{\centering}|m{0.7cm}<{\centering}|m{0.68cm}<{\centering}|m{0.76cm}<{\centering}|m{0.78cm}<{\centering}|m{0.74cm}<{\centering}|
m{0.74cm}<{\centering}|m{0.74cm}<{\centering}|m{0.88cm}<{\centering}|m{0.88cm}<{\centering}|m{0.74cm}<{\centering}|m{0.74cm}<{\centering}|m{0.74cm}<{\centering}|m{0.68cm}<{\centering}|}
\hline
\begin{minipage}{0.68cm}\vspace{1mm} \centering Index \vspace{1mm} \end{minipage}         &\begin{minipage}{0.68cm}\vspace{1mm} \centering 0 \vspace{1mm} \end{minipage}     &\begin{minipage}{0.7cm}\vspace{1mm} \centering 1 \vspace{1mm} \end{minipage}     &\begin{minipage}{0.68cm}\vspace{1mm} \centering 2 \vspace{1mm} \end{minipage}      &\begin{minipage}{0.76cm}\vspace{1mm} \centering 3 \vspace{1mm} \end{minipage}     &\begin{minipage}{0.78cm}\vspace{1mm} \centering 4 \vspace{1mm} \end{minipage}   &\begin{minipage}{0.74cm}\vspace{1mm} \centering 5 \vspace{1mm} \end{minipage}  &\begin{minipage}{0.74cm}\vspace{1mm} \centering 6 \vspace{1mm} \end{minipage}  &\begin{minipage}{0.74cm}\vspace{1mm} \centering 7 \vspace{1mm} \end{minipage}  &\begin{minipage}{0.88cm}\vspace{1mm} \centering 8 \vspace{1mm} \end{minipage}  &\begin{minipage}{0.88cm}\vspace{1mm} \centering 9 \vspace{1mm} \end{minipage}  &\begin{minipage}{0.74cm}\vspace{1mm} \centering 10 \vspace{1mm} \end{minipage}  &\begin{minipage}{0.74cm}\vspace{1mm} \centering 11 \vspace{1mm} \end{minipage}  &\begin{minipage}{0.74cm}\vspace{1mm} \centering 12 \vspace{1mm} \end{minipage}  &\begin{minipage}{0.68cm}\vspace{1mm} \centering 13 \vspace{1mm} \end{minipage}   \\
\hline
\begin{minipage}{0.68cm}\vspace{1mm} \centering Sound Class \vspace{1mm} \end{minipage}  &Alarm &\begin{minipage}{0.70cm}\vspace{1mm} \centering Crying Baby \vspace{1mm} \end{minipage} &\begin{minipage}{0.68cm}\vspace{1mm} \centering Crash \vspace{1mm} \end{minipage} &\begin{minipage}{0.76cm}\vspace{1mm} \centering Barking Dog \vspace{1mm} \end{minipage} &\begin{minipage}{0.78cm}\vspace{1mm} \centering Running Engine \vspace{1mm} \end{minipage} &Female Scream &Female Speech &\begin{minipage}{0.74cm}\vspace{1mm} \centering Burning Fire \vspace{1mm} \end{minipage} &Footsteps &Knocking Door &Male Scream  &Male Speech  &Ringing Phone  &Piano     \\
\hline
\end{tabular}
\end{table*}

We extract two types of features for each of the two datasets, FOA and MIC. Using STFT with a Hamming window of length 1024 samples and a 50\% overlap, a linear spectrogram for each channel is extracted. Then 64-dimensional log Mel-spectrogram feature vector is extracted for both datasets. The second type of feature is format-specific. For FOA dataset, acoustic intensity vector (IV) computed at each of the 64 Mel-bands is extracted while for the MIC dataset generalized cross-correlation phase transform (GCC-PHAT) is extracted similar to \cite{Cao2019,cao2019polyphonic} with one dimension equal to the number of Mel-band filters. Both IV and GCC-PHAT are then stacked with log Mel-spectrogram to serve as the input features. IV is extracted for FOA format data and can be calculated as ${\boldsymbol I} = p{\mathbf v}$. Variable $p$ denotes sound pressure and can be obtained with channel $W$ of the FOA signal, which corresponds to an omnidirectional microphone. Variable ${\mathbf v} = [{\mathbf v}_Y,{\mathbf v}_Z,{\mathbf v}_X]^T$ denotes the particle velocity vector and can be estimated using channels $Y$, $Z$ and $X$ of the FOA signal. Therefore, there are 3 channels of IV features, hence up to 7 feature maps for FOA signals along with 4 channels of log Mel-spectrogram features. For MIC signals, there are 4 channels of log Mel-spectrogram features and GCC-PHAT features are computed for each pair of the 4 channels. Finally, 6 channels of GCC-PHAT features are extracted, hence up to 10 feature maps. We use both FOA and MIC datasets, so 17 input feature maps are used to train the models.

The TFM augmentation approach is applied to each acoustic feature in each batch. For every acoustic feature, we multiply masks on time and frequency dimensions for the first 11 feature maps. The last 6 feature maps containing DOA information are not applied with the masks. The parameters of TFM are inspired by the SpecAugment proposed in ASR \cite{park2019specaugment}. Frequency masking parameter is set to at most 50\% of the dimension and time masking parameter is a little smaller. Considering that a sample has a long length of 60 seconds, we perform time masking every 2 seconds. Specifically, time mask length is randomly selected from zero to 35 frames, and masking is applied every 100 frames. Frequency mask length is randomly selected from zero to 30 bins.

A joint measurement on the performances of localization and detection of sound events is performed as suggested in \cite{mesaros2019joint}. Location-dependent detection metrics that count correct and erroneous detections within certain spatial error allowances and classification-dependent localization metrics that measure the spatial error between sound events with the same label are used to evaluate the SED and SSL performances, respectively. To compute SED metrics, some intermediate statistics, such as true positive ($\emph{TP}$), false positive ($\emph{FP}$ or insertion error $\emph{I}$), false negative ($\emph{FN}$ or deletion error $\emph{D}$), and substitution error $\emph{S}$, need to be counted first. Considering that a $\emph{TP}$ is predicted only when the spatial error for the detected event is within the given threshold of $20^{\text{o}}$ from the reference, two location-dependent detection metrics, error rate ($\emph{ER}_{20^{\text{o}}}$) and F-score ($\emph{F}_{20^{\text{o}}}$), are then calculated as follows:
\begin{equation}
P=\frac{TP}{TP+FP}, R=\frac{TP}{TP+FN} ,
\end{equation}
\begin{equation}
\emph{ER}_{20^{\text{o}}}=\frac{D+I+S}{N}, \emph{F}_{20^{\text{o}}}=\frac{2PR}{P+R} ,
\end{equation}
where $\emph{N}$ is the total number of reference sound events. $P$ and $R$ denote the precision and recall metrics, respectively.

Classification-dependent localization metrics are computed only across each class, instead of across all outputs. The first is the localization error $\emph{LE}_{\text{CD}}$ which expresses the average angular distance between predictions and references of the same class and can be calculated as
\begin{equation}
\emph{LE}_{\text{CD}}=\rm{arccos}({\rm\mathbf u}_{\rm ref} \cdot {\rm\mathbf u}_{\rm pre}) ,
\end{equation}
where ${\rm\mathbf u}_{\rm ref}$ and ${\rm\mathbf u}_{\rm pre}$ denote the unit Cartesian position vectors of reference sound event and predicted sound event, respectively. The subscript refers to classification-dependent. The second is a simple localization recall metric $\emph{LR}_{\text{CD}}$ which expresses the true positive rate of how many of these localization estimates are detected in a class out of the total class instances.

All metrics are computed in one-second non-overlapping segments to alleviate the effect of onset/offset subjectivity in reference annotations. With these four metrics, an early stopping SELD score $\emph{SELD}_{\rm score}$ can be computed as follows
\begin{small}
\begin{equation}
\emph{SELD}_{\rm score}=\frac{\emph{ER}_{20^{\text{o}}}+(1-\emph{F}_{20^{\text{o}}})+\emph{LE}_{\text{CD}}^{'}+(1-\emph{LR}_{\text{CD}})}{4} ,
\end{equation}
\end{small}
where $\emph{LE}_{\text{CD}}^{'}=\emph{LE}_{\text{CD}}/\pi$. The $\emph{SELD}_{\rm score}$ is an overall performance metric for the SELD task. The model with the smallest $\emph{SELD}_{\rm score}$ on the validation split is chosen as the best model.

Audio clips with a length of 60 seconds are used for training all model architectures with an Adam optimizer \cite{kingma2014adam}. The learning rate is set to 0.001 and is decreased by 50\% if the SELD score of the validation split does not improve in 80 consecutive epochs. A threshold of 0.5 is used to assess the predicted results of the SELD model. For the Conformer module, the number of attention heads $h$ is set to 8, and the dimension of input vector $d$ is set to 512. For simplicity, $d_k$ and $d_v$ are both equal to 64. We use a kernel size of 51 for the depthwise convolution. The module number $B$ shown in Fig.~\ref{fig:conformer} is set to 8. All experiments in this study were performed using the PyTorch toolkit.

\subsection{Results Based on Different Acoustic Models}
\label{subsec:diff_dnn}

First in Table~\ref{DNNs}, we compare different acoustic models on SELD without using any data augmentation. The official baseline SELDnet system \cite{politis2020dataset} was compared with the best ResNet-GRU model proposed in our submitted system for the DCASE 2020 Challenge \cite{Du2020_task3_report,wang2021model}. In addition, the performance of ResNet-Conformer is also compared.
\begin{table}\scriptsize
\caption{A performance comparison for different models on the development set without data augmentation.}
\label{DNNs}
\centering
\begin{tabular}{c c c c c c}
\toprule
System  &$\emph{ER}_{20^{\text{o}}}$ &$\emph{F}_{20^{\text{o}}}$ &$\emph{LE}_{\text{CD}}$ &$\emph{LR}_{\text{CD}}$  &$\emph{SELD}_{\rm score}$\\
\midrule
Baseline-MIC          &0.78          &31.4\%          &$27.3^{\rm{o}}$              &59.0\%   &0.51        \\
\midrule
ResNet-GRU-MIC        &0.67          &42.1\%          &$23.7^{\rm{o}}$              &67.0\%   &0.43        \\
\midrule
Baseline-FOA          &0.72          &37.4\%          &$22.8^{\rm{o}}$              &60.7\%   &0.47        \\
\midrule
ResNet-GRU-FOA        &0.65          &46.1\%          &$19.2^{\rm{o}}$              &64.8\%   &0.41        \\
\midrule
ResNet-GRU-Both            &0.63          &47.6\%          &$18.7^{\rm{o}}$              &67.7\%   &0.40        \\
\midrule
ResNet-Conformer-Both      &0.51          &58.6\%          &$15.6^{\rm{o}}$              &73.3\%   &0.32        \\
\bottomrule
\end{tabular}
\end{table}

The first two rows represent the official baseline and our ResNet-GRU systems trained with only the MIC data set. ``Baseline-FOA'' and ``ResNet-GRU-FOA'' are compared in the third and fourth rows of Table~\ref{DNNs} using the FOA data. The proposed ResNet-GRU architecture outperforms the baseline SELDnet for both MIC and FOA formats on all four evaluation metrics. The main reason may be that the use of residual connection helps the models to capture more useful shift-invariant local features from the input acoustic features. ``ResNet-GRU-Both'' shown in the bottom row is trained with the concatenated features extracted from both MIC and FOA data sets, yielding high scores than the models trained with separated features. By replacing the GRU module with the Conformer module, ``ResNet-Conformer-Both'' achieves consistent improvements for both SED and SSL metrics over ``ResNet-GRU-Both'', which demonstrates that the Conformer is more effective in modeling context dependencies than GRU. Compared to the two official baseline systems, the best ResNet-Conformer model achieves 37.2\% and 31.9\% relative improvements on the SELD scores, respectively.
\begin{figure}
  \centering

  \begin{minipage}[b]{0.49\linewidth}
  \centering
  \includegraphics[width=0.92\linewidth]{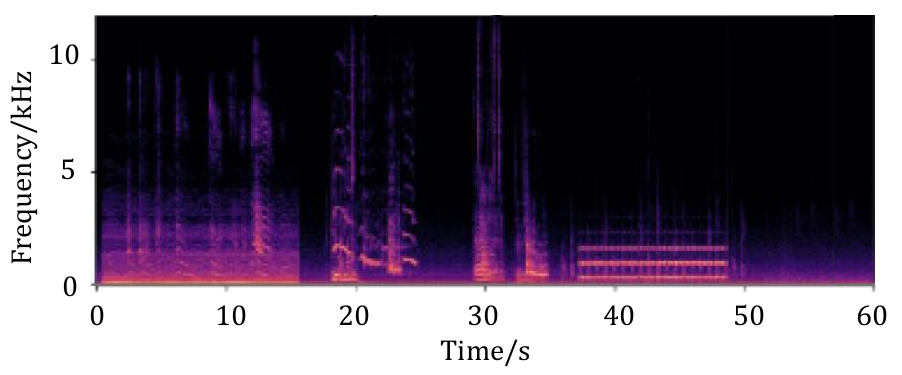}
  \centerline{\tiny{(a) Spectrogram}}
  \end{minipage}
  \label{fig:spec}
  \hfill
  \begin{minipage}[b]{0.49\linewidth}
  \centering
  \includegraphics[width=0.92\linewidth]{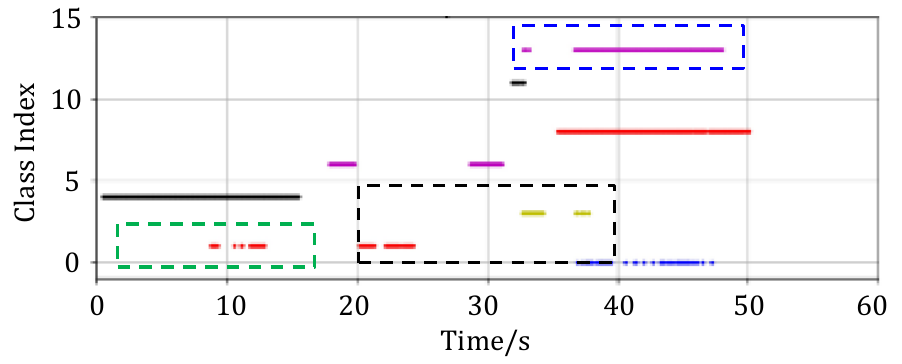}
  \centerline{\tiny{(b) SED prediction of ResNet-GRU}}
  \end{minipage}
  \label{fig:ref}
  \hfill
  \begin{minipage}[b]{0.49\linewidth}
  \centering
  \includegraphics[width=0.92\linewidth]{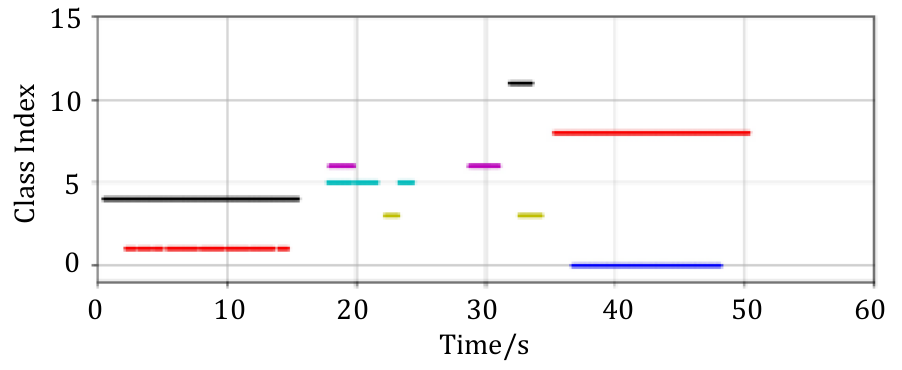}
  \centerline{\tiny{(c) SED Reference}}
  \end{minipage}
  \label{fig:out1}
  \hfill
  \begin{minipage}[b]{0.49\linewidth}
  \centering
  \includegraphics[width=0.92\linewidth]{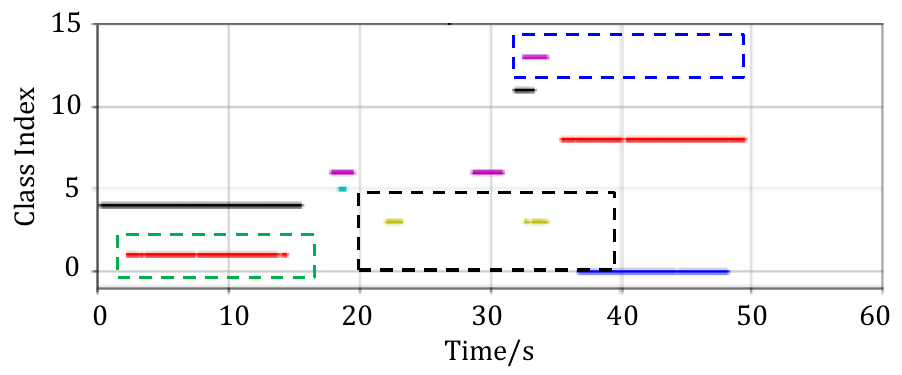}
  \centerline{\tiny{(d) SED prediction of ResNet-Conformer}}
  \end{minipage}
  \label{fig:out2}

  \centering
  \caption{An example comparison of ResNet-GRU and ResNet-Conformer.}
  \label{fig:example}
\end{figure}

Fig.~\ref{fig:example} illustrates a visualization of the SED prediction of the class index of sound events as listed in Table~\ref{class} using ResNet-GRU and ResNet-Conformer without data augmentation. The SED-predicted indices of ResNet-Conformer are more accurate than those of ResNet-GRU. In the first 20 seconds, two sound events, i.e., ``Crying Baby'' (with index 1) and ``Running Engine'' (with index 4), occur at the same time. ResNet-GRU could not well detect the ``Crying Baby'' event segments, but for ResNet-Conformer these segments are correctly predicted as shown in the green dashed rectangular boxes, which proves the effectiveness of the Conformer networks. As shown in Fig.~\ref{fig:example}(a) and Fig.~\ref{fig:example}(c), the ``Barking Dog''  (with index 3) event happens twice and there is a period between them. However, ResNet-GRU only recognizes the second instance and wrongly predicts the first occurrence of ``Barking Dog'' as ``Crying Baby''. ResNet-Conformer detects the two occurrences as shown in the black dashed rectangular box of Fig.~\ref{fig:example}(d), indicating its superiority in modeling short segments. From about 35 to 50 seconds, there exist two sound events, ``Alarm''  (with index 0) and ``Footsteps''  (with index 8). ResNet-Conformer wrongly recognizes it as ``Piano'' at the beginning, but when a longer sequence could be observed it corrects the error as shown in the blue dashed rectangular box of Fig.~\ref{fig:example}(d) and correctly predicts the two sound events. ResNet-GRU, however, wrongly recognizes it as ``Piano'' even after a long duration and predicts it as three separate sound events. This example shows Conformer's superiority over GRU. By using Conformer, it is more likely than GRU to capture both local and global context dependencies.

\subsection{Results Based on ACS Spatial Augmentation}
In \cite{MazzonYasuda2019}, the authors proposed an augmentation method using the FOA data set, containing sixteen patterns of spatial augmentation whereas our proposed ACS approach can be applied to not only the FOA but also MIC data sets. A performance comparison between these two methods is shown in Table~\ref{result1}. To make a fair comparison, we apply the method in \cite{MazzonYasuda2019} to augment the same amount of training data as our proposed ACS approach. ``ACS-FOA'' denotes the system trained only with the FOA set while ``ACS'' denotes the system trained with both FOA and MIC sets. The difference of ``ACS-FOA'' from \cite{MazzonYasuda2019} is that only eight patterns/transformations in Table~\ref{ACS} were adopted in our approach. It is noted that similar results are obtained by these two systems. This indicates that the eight patterns adopted by ACS already contain enough useful DOA information, and adding the other eight patterns may lead to information redundancy since they just apply reflections with respect to the $xy:z=0$ plane when compared with the eight patterns adopted by ACS. By comparing the bottom two rows in Table~\ref{result1}, we can see that ACS outperforms ACS-FOA when applying spatial augmentation to the MIC set. This result verifies the usefulness of feature fusion of both FOA and MIC data sets, which is also shown in Section~\ref{sec:exp}-A.
\begin{table}
\caption{A performance comparison of different ACS approaches using the ResNet-GRU model.}
\label{result1}
\centering
\begin{tabular}{c c c c c c}
\toprule
System  &$\emph{ER}_{20^{\text{o}}}$ &$\emph{F}_{20^{\text{o}}}$ &$\emph{LE}_{\text{CD}}$ &$\emph{LR}_{\text{CD}}$ &$\emph{SELD}_{\rm score}$\\
\midrule
\cite{MazzonYasuda2019}     &0.44          &63.9\%          &$14.7^{\rm{o}}$              &75.3\%     &0.28      \\
\midrule
ACS-FOA      &0.44          &64.5\%          &$13.1^{\rm{o}}$              &73.7\%     &0.28      \\
\midrule
ACS          &0.42          &67.0\%          &$12.4^{\rm{o}}$              &75.6\%     &0.27      \\
\bottomrule
\end{tabular}
\end{table}

\subsection{Results Based on Individual Data Augmentation}
We next adopt ResNet-GRU as the acoustic model to compare the system performances, as listed in Table~\ref{result2}, when applying four data augmentation approaches, namely ACS, MCS, TDM, and TFM, individually. All four augmentation approaches generate a similar size of training data for about 55 hours.

\begin{table}
\caption{A performance comparison when applying four augmentation approaches individually using the ResNet-GRU model.}
\label{result2}
\centering
\begin{tabular}{c c c c c c}
\toprule
System  &$\emph{ER}_{20^{\text{o}}}$ &$\emph{F}_{20^{\text{o}}}$ &$\emph{LE}_{\text{CD}}$ &$\emph{LR}_{\text{CD}}$ &$\emph{SELD}_{\rm score}$\\
\midrule
ResNet-GRU &0.63          &47.6\%          &$18.7^{\rm{o}}$              &67.7\%          &0.40     \\
\midrule
ACS          &0.42          &67.0\%          &$12.4^{\rm{o}}$              &75.6\%     &0.27      \\
\midrule
MCS          &0.44          &65.2\%          &$13.6^{\rm{o}}$              &76.3\%     &0.28      \\
\midrule
TDM          &0.50          &60.5\%          &$14.0^{\rm{o}}$              &72.3\%     &0.31      \\
\midrule
TFM          &0.57          &51.5\%          &$18.8^{\rm{o}}$              &72.7\%     &0.36      \\
\bottomrule
\end{tabular}
\end{table}

We can make the following observations: (i) all four metrics yield gains except for the $\emph{LE}_{\text{CD}}$ metric of the TFM approach. Since only several hours of audio data is available, applying masks on log Mel-spectrogram features may not bring performance gain to the SSL task, which has also been observed in \cite{zhang2019data}; (ii) the ACS and MCS approaches achieve consistent improvements for both SED and SSL metrics. This demonstrates that increasing the diversity of DOA representations is very effective for the SELD task. The ACS approach can be applied to complete signals in the development data set, while the MCS approach can only be applied to audio segments containing non-overlapping and non-moving sound events. Thus the overall performance $\emph{SELD}_{\rm score}$ of the ACS approach is slightly better than that of the MCS approach; and (iii) for the TDM approach, both SED and SSL metrics improve even though no new DOA presentation is generated, indicating mixing two non-overlapping sound signals in the time domain helps model robustness to unseen samples. In summary for the ResNet-GRU system with no data augmentation, ACS, MCS, TDM, and TFM individually yield 32.5\%, 30.0\%, 22.5\%, and 10.0\% relative SELD score reductions, respectively.

\subsection{Results Based on Four-Stage Data Augmentation}

We next evaluate the system performances when using the four augmentation techniques. The four-stage data augmentation scheme was used to exploit the complementarity among the four approaches and our submitted ResNet-GRU ensemble system ranks the first place for the SELD task of the DCASE 2020 Challenge \cite{DCASE2020_task3_report}. Since ACS can be applied to the whole development data set, we perform ACS on the original data in the first stage. MCS aims to simulate new DOA presentations for audio segments containing isolated and static sound events, on which the TDM approach can be applied. So MCS is performed in the second stage and TDM is performed in the following third stage. With a larger data set now, we apply TFM in the final stage.

Table~\ref{result3} lists performance comparisons when applying the four-stage data augmentation scheme using ResNet-GRU and ResNet-Conformer, respectively. The first and second columns denote the systems and the corresponding training data size, respectively. ACS is performed on the original data in the first stage, generating a 55-hour training set. Then we apply MCS to the 55-hour set, generating a larger 155-hour set. TDM and TFM are conducted similarly, and finally a 255-hour training set is obtained. For ResNet-GRU, each augmentation approach achieves performance gains for the SED and SSL metrics. When applying ACS on the original data set, ``S2'' achieves a SELD score of 0.27 lower than 0.40 for S1 without any augmentation. When applying MCS, TDM, and TFM separately on the original data set used in S1, the SELD scores are worse than the ACS approach as shown in Table~\ref{result2}. However, by using the proposed four-stage data augmentation scheme, consistent performance gains are yielded in ``S3'', ``S4'', and ``S5''. Compared to the model without using data augmentation in S1, these four systems achieve 32.5\%, 40.0\%, 45.0\%, and 55.0\% relative  SELD score reductions, respectively. The training time of a model is proportional to the size of the data. If the four-stage data augmentation scheme is used, it takes 1$\sim$2 days to finish training ``S5'', which is relatively acceptable and worthwhile compared with the performance of S1.
\begin{table}
\caption{A performance comparison by combining four augmentation approaches. (S1:ResNet-GRU, S2:S1+ACS, S3:S2+MCS, S4:S3+TDM, S5:S4+TFM, S6:ResNet-Conformer+ACS+MCS+TDM+TFM)}
\label{result3}
\centering
\begin{tabular}{c c c c c c c}
\toprule
System & Size (h) &$\emph{ER}_{20^{\text{o}}}$ &$\emph{F}_{20^{\text{o}}}$ &$\emph{LE}_{\text{CD}}$ &$\emph{LR}_{\text{CD}}$  &$\emph{SELD}_{\rm score}$\\
\midrule
 S1  &8     &0.63          &47.6\%          &$18.7^{\rm{o}}$              &67.7\%          &0.40      \\
\midrule
S2   &55    &0.42          &67.0\%          &$12.4^{\rm{o}}$              &75.6\%     &0.27   \\
\midrule
S3  &155   &0.37          &70.9\%          &$10.2^{\rm{o}}$              &77.5\%     &0.24    \\
\midrule
S4  &255   &0.34          &73.2\%          &$9.8^{\rm{o}}$               &79.3\%     &0.22    \\
\midrule
S5  &255  &$\textbf{0.27}$          &$\textbf{78.1}$\%         &$\textbf{8.5}^{\rm{o}}$               &$\textbf{83.6}$\%     &$\textbf{0.18}$    \\
\midrule
S6  &255  &$\textbf{0.26}$          &$\textbf{80.0}$\%          &$\textbf{8.0}^{\rm{o}}$               &$\textbf{84.3}$\%     &$\textbf{0.17}$    \\
\bottomrule
\end{tabular}
\end{table}

Next the four-stage data augmentation approach is also evaluated using the proposed ResNet-Conformer model as shown in the last row of Table~\ref{result3}. The SELD score without using data augmentation for ResNet-Conformer is 0.32 as listed in Table~\ref{DNNs}, yielding a 20\% relative reduction from 0.40 for ResNet-GRU in S1. When comparing performances between ResNet-Conformer and ResNet-GRU, the gains are reduced with the proposed data augmentation approaches. The results in the bottom rows highlighted in bold fonts in Table~\ref{result3} after applying all four techniques show only slight performance differences. Clearly for deep learning, the four-stage scheme can largely increase the data diversity, thus improving the generalization ability of acoustic models.

\begin{figure}
  \centering

  \begin{minipage}[b]{0.49\linewidth}
  \centering
  \includegraphics[width=0.92\linewidth]{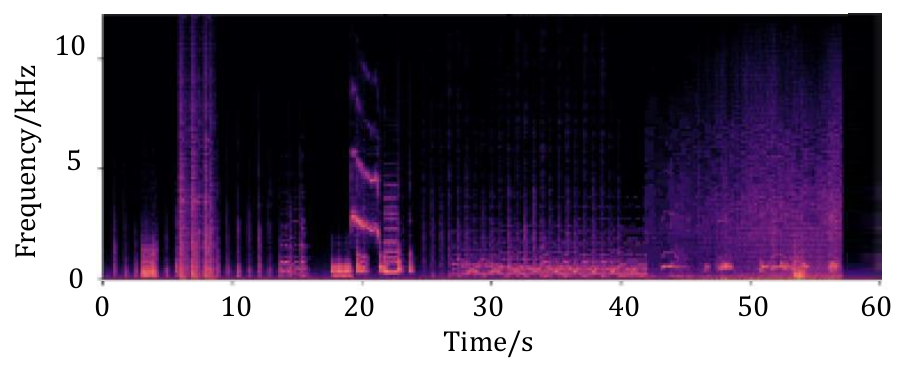}
  \centerline{\tiny{(a) Spectrogram}}
  \end{minipage}
  \label{fig:spec}
  \hfill
  \begin{minipage}[b]{0.49\linewidth}
  \centering
  \includegraphics[width=0.92\linewidth]{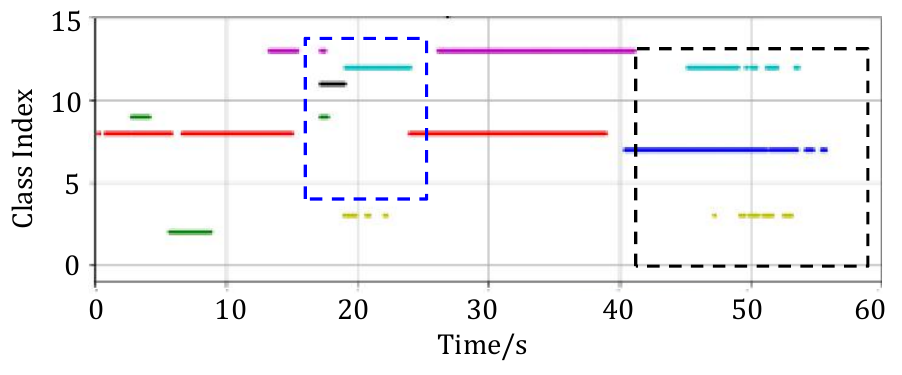}
  \centerline{\tiny{(b) SED prediction without data augmentation}}
  \end{minipage}
  \label{fig:ref}
  \hfill
  \begin{minipage}[b]{0.49\linewidth}
  \centering
  \includegraphics[width=0.92\linewidth]{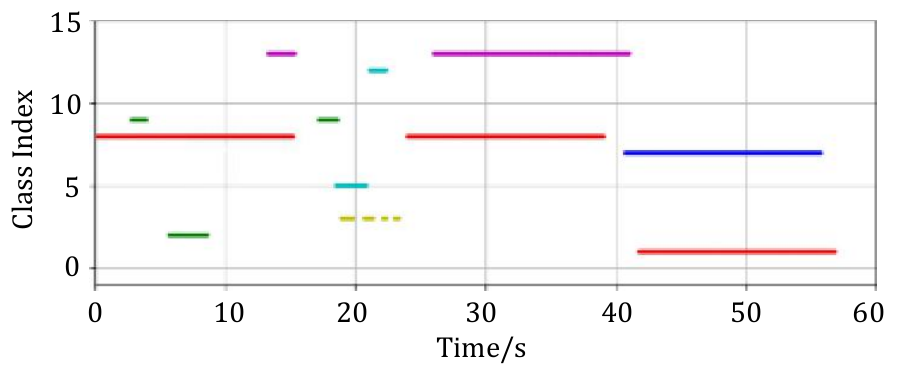}
  \centerline{\tiny{(c) SED Reference}}
  \end{minipage}
  \label{fig:out1}
  \hfill
  \begin{minipage}[b]{0.49\linewidth}
  \centering
  \includegraphics[width=0.92\linewidth]{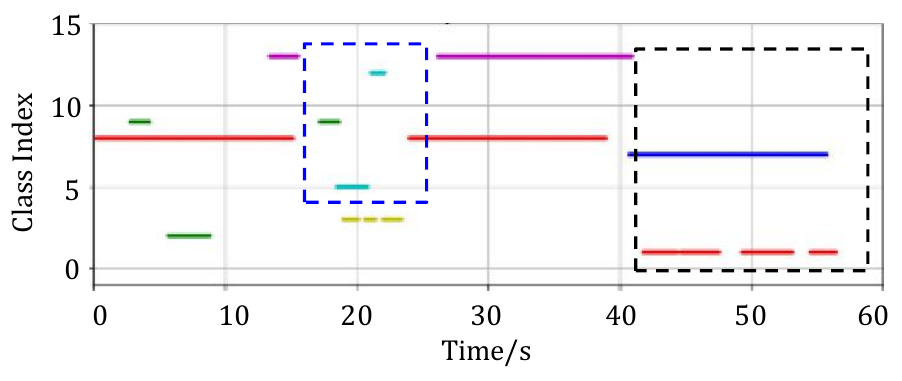}
  \centerline{\tiny{(d) SED prediction with data augmentation}}
  \end{minipage}
  \label{fig:out2}

  \centering
  \caption{An example comparison of ResNet-Conformer with or without four-stage data augmentation.}
  \label{fig:example2}
\end{figure}

Fig.~\ref{fig:example2} shows an example of the SED prediction using ResNet-Conformer model with and without four-stage data augmentation. For the two segments from beginning to 15 seconds and from 25 to 40 seconds, ResNet-Conformer predicts correct results both with and without data augmentation. As shown in the blue dashed rectangular boxes, when data augmentation is not used, the model tends to wrongly predict the shorter sound events. But with data augmentation, the model can output correct predictions. In the last 20 seconds, the model trained without data augmentation cannot recognize overlapping sound events, and instead predicts two wrong events as shown in the black dashed rectangular box of Fig.~\ref{fig:example2}(b). However, the model correctly predicts the overlapping sound events when adopting the proposed four-stage data augmentation approach.

\subsection{Results Based on DCASE 2022 Challenge}

Finally, we list the experimental results of SELD for DCASE 2022 Challenge in Table~\ref{DCASE2022}. The first two rows represent the official baseline and our ResNet-Conformer systems, both trained with the FOA data set. The ResNet-Conformer system adopted in DCASE 2022 Challenge is trained with fewer parameters. The ``ResNet-Conformer+FSDA'' denotes the ResNet-Conformer system when applying the four-stage data augmentation approach. As shown in Table~\ref{DCASE2022}, consistent improvements are achieved by using the proposed acoustic model and data augmentation method. The last row represents our submitted system with model ensemble strategies, which ranks the first place in DCASE 2022 Challenge \cite{Du_NERCSLIP_task3_report,DCASE2022_task3_report}.

\begin{table}\scriptsize
\caption{A performance comparison of different systems on the FOA data set for DCASE 2022 Challenge.}
\label{DCASE2022}
\centering
\begin{tabular}{c c c c c c}
\toprule
System  &$\emph{ER}_{20^{\text{o}}}$ &$\emph{F}_{20^{\text{o}}}$ &$\emph{LE}_{\text{CD}}$ &$\emph{LR}_{\text{CD}}$  &$\emph{SELD}_{\rm score}$\\
\midrule
Baseline         &0.71         &21\%          &$29.3^{\rm{o}}$              &46\%   &0.55        \\
\midrule
ResNet-Conformer        &0.65          &33\%          &$23.5^{\rm{o}}$              &58\%   &0.47       \\
\midrule
ResNet-Conformer+FSDA          &0.49          &56\%          &16.0$^{\rm{o}}$              &72\%   &0.32        \\
\midrule
Submission        &0.41         &64\%          &$14.9^{\rm{o}}$              &73\%   &0.28       \\
\bottomrule
\end{tabular}
\end{table}

\section{Conclusion}
\label{sec:conc}
This study focuses on data augmentation and acoustic modeling for the SELD task. Two novel spatial augmentation approaches, namely ACS and MCS, are proposed to deal with data sparsity for deep learning based acoustic modeling. The ACS approach can be applied to complete signals while the MCS approach is suitable for audio segments containing isolated and static sound events, both of which aim at increasing the diversity of DOA representations. We adopt a four-stage data augmentation scheme to improve the performance step-by-step. We also employ a Conformer architecture that combines convolution and Transformer to model both global and local context dependencies in an audio sequence and propose a ResNet-Conformer architecture. Tested on the data sets of the DCASE 2020 and 2022 Challenges, our proposed data augmentation approaches and ResNet-Conformer model have greatly improved the performance, ranking our submitted systems in the first place in the SELD task of both DCASE 2020 and 2022 Challenges.

\section*{Acknowledgment}

The authors would like to thank Yuxuan Wang, Tairan Chen, Zijun Jing and Yi Fang for their help on some experiments.

\ifCLASSOPTIONcaptionsoff
  \newpage
\fi


\scriptsize
\bibliographystyle{IEEEtran}

\bibliography{IEEEabrv,mybibfile}

\begin{thebibliography}{10}
\providecommand{\url}[1]{#1}
\csname url@samestyle\endcsname
\providecommand{\newblock}{\relax}
\providecommand{\bibinfo}[2]{#2}
\providecommand{\BIBentrySTDinterwordspacing}{\spaceskip=0pt\relax}
\providecommand{\BIBentryALTinterwordstretchfactor}{4}
\providecommand{\BIBentryALTinterwordspacing}{\spaceskip=\fontdimen2\font plus
\BIBentryALTinterwordstretchfactor\fontdimen3\font minus
  \fontdimen4\font\relax}
\providecommand{\BIBforeignlanguage}[2]{{%
\expandafter\ifx\csname l@#1\endcsname\relax
\typeout{** WARNING: IEEEtran.bst: No hyphenation pattern has been}%
\typeout{** loaded for the language `#1'. Using the pattern for}%
\typeout{** the default language instead.}%
\else
\language=\csname l@#1\endcsname
\fi
#2}}
\providecommand{\BIBdecl}{\relax}
\BIBdecl
\renewcommand{\BIBentryALTinterwordstretchfactor}{4}

\bibitem{wang1997voice}
H.~Wang and P.~Chu, ``Voice source localization for automatic camera pointing
  system in videoconferencing,'' in \emph{Proc. {IEEE} Int. Conf. Acoust.,
  Speech Signal Process.}, vol.~1, 1997, pp. 187--190.

\bibitem{swietojanski2014convolutional}
P.~Swietojanski, A.~Ghoshal, and S.~Renals, ``Convolutional neural networks for
  distant speech recognition,'' \emph{{IEEE} Signal Process. Lett.}, vol.~21,
  no.~9, pp. 1120--1124, 2014.

\bibitem{valenzise2007scream}
G.~Valenzise, L.~Gerosa, M.~Tagliasacchi, F.~Antonacci, and A.~Sarti, ``Scream
  and gunshot detection and localization for audio-surveillance systems,'' in
  \emph{Proc. {IEEE} Conf. Adv. Video Signal Based Surveillance}, 2007, pp.
  21--26.

\bibitem{foggia2015audio}
P.~Foggia, N.~Petkov, A.~Saggese, N.~Strisciuglio, and M.~Vento, ``Audio
  surveillance of roads: A system for detecting anomalous sounds,''
  \emph{{IEEE} Trans. Intell. Transp. Syst.}, vol.~17, no.~1, pp. 279--288,
  2015.

\bibitem{heittola2010audio}
T.~Heittola, A.~Mesaros, A.~Eronen, and T.~Virtanen, ``Audio context
  recognition using audio event histograms,'' in \emph{Proc. {IEEE} Eur. Signal
  Process. Conf.}, 2010, pp. 1272--1276.

\bibitem{mesaros2010acoustic}
A.~Mesaros, T.~Heittola, A.~Eronen, and T.~Virtanen, ``Acoustic event detection
  in real life recordings,'' in \emph{Proc. {IEEE} Eur. Signal Process. Conf.},
  2010, pp. 1267--1271.

\bibitem{heittola2013context}
T.~Heittola, A.~Mesaros, A.~Eronen, and T.~Virtanen, ``Context-dependent sound
  event detection,'' \emph{EURASIP J. Audio, Speech, Music Process.}, vol.
  2013, no.~1, p.~1, 2013.

\bibitem{gemmeke2013exemplar}
J.~F. Gemmeke, L.~Vuegen, P.~Karsmakers, B.~Vanrumste \emph{et~al.}, ``An
  exemplar-based {NMF} approach to audio event detection,'' in \emph{Proc. IEEE
  Workshop Appl. Signal Process. Audio Acoust.}, 2013, pp. 1--4.

\bibitem{mesaros2015sound}
A.~Mesaros, T.~Heittola, O.~Dikmen, and T.~Virtanen, ``Sound event detection in
  real life recordings using coupled matrix factorization of spectral
  representations and class activity annotations,'' in \emph{Proc. IEEE Int.
  Conf. Acoust., Speech Signal Process.}, 2015, pp. 151--155.

\bibitem{mcloughlin2015robust}
I.~McLoughlin, H.~Zhang, Z.~Xie, Y.~Song, and W.~Xiao, ``Robust sound event
  classification using deep neural networks,'' \emph{IEEE/ACM Trans. Audio,
  Speech, Lang. Process.}, vol.~23, no.~3, pp. 540--552, 2015.

\bibitem{piczak2015environmental}
K.~J. Piczak, ``Environmental sound classification with convolutional neural
  networks,'' in \emph{Proc. {IEEE} Int. Workshop Mach. Learning Signal
  Process.}, 2015, pp. 1--6.

\bibitem{zhang2015robust}
H.~Zhang, I.~McLoughlin, and Y.~Song, ``Robust sound event recognition using
  convolutional neural networks,'' in \emph{Proc. {IEEE} Int. Conf. Acoust.,
  Speech Signal Process.}, 2015, pp. 559--563.

\bibitem{phan2016robust}
H.~Phan, L.~Hertel, M.~Maass, and A.~Mertins, ``Robust audio event recognition
  with 1-max pooling convolutional neural networks,'' in \emph{Proc.
  Interspeech}, 2016, pp. 3653--3657.

\bibitem{wang2016audio}
Y.~Wang, L.~Neves, and F.~Metze, ``Audio-based multimedia event detection using
  deep recurrent neural networks,'' in \emph{Proc. {IEEE} Int. Conf. Acoust.,
  Speech Signal Process.}, 2016, pp. 2742--2746.

\bibitem{parascandolo2016recurrent}
G.~Parascandolo, H.~Huttunen, and T.~Virtanen, ``Recurrent neural networks for
  polyphonic sound event detection in real life recordings,'' in \emph{Proc.
  {IEEE} Int. Conf. Acoust., Speech Signal Process.}, 2016, pp. 6440--6444.

\bibitem{hayashi2017duration}
T.~Hayashi, S.~Watanabe, T.~Toda, T.~Hori, J.~Le~Roux, and K.~Takeda,
  ``Duration-controlled {LSTM} for polyphonic sound event detection,''
  \emph{IEEE/ACM Trans. Audio, Speech, Lang. Process.}, vol.~25, no.~11, pp.
  2059--2070, 2017.

\bibitem{sabour2017dynamic}
S.~Sabour, N.~Frosst, and G.~E. Hinton, ``Dynamic routing between capsules,''
  in \emph{Proc. Adv. Neural Inf. Process. Syst.}, 2017, pp. 3856--3866.

\bibitem{vesperini2019polyphonic}
F.~Vesperini, L.~Gabrielli, E.~Principi, and S.~Squartini, ``Polyphonic sound
  event detection by using capsule neural networks,'' \emph{IEEE J. Sel. Topics
  Signal Process.}, vol.~13, no.~2, pp. 310--322, 2019.

\bibitem{liu2018capsule}
Y.~Liu, J.~Tang, Y.~Song, and L.~Dai, ``A capsule based approach for polyphonic
  sound event detection,'' in \emph{Proc. Asia-Pacific Signal Inf. Process.
  Assoc.}, 2018, pp. 1853--1857.

\bibitem{cakir2017convolutional}
E.~Cak{\i}r, G.~Parascandolo, T.~Heittola, H.~Huttunen, and T.~Virtanen,
  ``Convolutional recurrent neural networks for polyphonic sound event
  detection,'' \emph{IEEE/ACM Trans. Audio, Speech, Lang. Process.}, vol.~25,
  no.~6, pp. 1291--1303, 2017.

\bibitem{adavanne2018sound}
S.~Adavanne, A.~Politis, J.~Nikunen, and T.~Virtanen, ``Sound event
  localization and detection of overlapping sources using convolutional
  recurrent neural networks,'' \emph{IEEE J. Sel. Topics Signal Process.},
  vol.~13, no.~1, pp. 34--48, 2018.

\bibitem{cao2019polyphonic}
Y.~Cao, Q.~Kong, T.~Iqbal, F.~An, W.~Wang, and M.~D. Plumbley, ``Polyphonic
  sound event detection and localization using a two-stage strategy,'' in
  \emph{Proc. Detection Classification Acoust. Scenes Events Workshop}, 2019,
  pp. 30--34.

\bibitem{schmidt1986multiple}
R.~Schmidt, ``Multiple emitter location and signal parameter estimation,''
  \emph{IEEE Trans. Antennas Propag.}, vol.~34, no.~3, pp. 276--280, 1986.

\bibitem{roy1989esprit}
R.~Roy and T.~Kailath, ``{ESPRIT}-estimation of signal parameters via
  rotational invariance techniques,'' \emph{IEEE Trans. Acoust., Speech, Signal
  Process.}, vol.~37, no.~7, pp. 984--995, 1989.

\bibitem{teutsch2006acoustic}
H.~Teutsch and W.~Kellermann, ``Acoustic source detection and localization
  based on wavefield decomposition using circular microphone arrays,'' \emph{J.
  Acoust. Soc. Amer.}, vol. 120, no.~5, pp. 2724--2736, 2006.

\bibitem{knapp1976generalized}
C.~Knapp and G.~Carter, ``The generalized correlation method for estimation of
  time delay,'' \emph{IEEE Trans. Acoust., Speech, Signal Process.}, vol.~24,
  no.~4, pp. 320--327, 1976.

\bibitem{brandstein1997robust}
M.~S. Brandstein and H.~F. Silverman, ``A robust method for speech signal
  time-delay estimation in reverberant rooms,'' in \emph{Proc. {IEEE} Int.
  Conf. Acoust., Speech Signal Process.}, vol.~1, 1997, pp. 375--378.

\bibitem{do2007real}
H.~Do, H.~F. Silverman, and Y.~Yu, ``A real-time {SRP-PHAT} source location
  implementation using stochastic region contraction ({SRC}) on a
  large-aperture microphone array,'' in \emph{Proc. {IEEE} Int. Conf. Acoust.,
  Speech Signal Process.}, vol.~1, 2007, pp. 121--124.

\bibitem{ferguson2018sound}
E.~L. Ferguson, S.~B. Williams, and C.~T. Jin, ``Sound source localization in a
  multipath environment using convolutional neural networks,'' in \emph{Proc.
  {IEEE} Int. Conf. Acoust., Speech Signal Process.}, 2018, pp. 2386--2390.

\bibitem{liu2018direction}
Z.-M. Liu, C.~Zhang, and S.~Y. Philip, ``Direction-of-arrival estimation based
  on deep neural networks with robustness to array imperfections,'' \emph{IEEE
  Trans. Antennas Propag.}, vol.~66, no.~12, pp. 7315--7327, 2018.

\bibitem{adavanne2018direction}
S.~Adavanne, A.~Politis, and T.~Virtanen, ``Direction of arrival estimation for
  multiple sound sources using convolutional recurrent neural network,'' in
  \emph{Proc. IEEE Eur. Signal Process. Conf.}, 2018, pp. 1462--1466.

\bibitem{pavlidi20153d}
D.~Pavlidi, S.~Delikaris-Manias, V.~Pulkki, and A.~Mouchtaris, ``3{D}
  localization of multiple sound sources with intensity vector estimates in
  single source zones,'' in \emph{Proc. IEEE Eur. Signal Process. Conf.}, 2015,
  pp. 1556--1560.

\bibitem{hafezi2017augmented}
S.~Hafezi, A.~H. Moore, and P.~A. Naylor, ``Augmented intensity vectors for
  direction of arrival estimation in the spherical harmonic domain,''
  \emph{IEEE/ACM Trans. Audio, Speech, Lang. Process.}, vol.~25, no.~10, pp.
  1956--1968, 2017.

\bibitem{yasuda2020sound}
M.~Yasuda, Y.~Koizumi, S.~Saito, H.~Uematsu, and K.~Imoto, ``Sound event
  localization based on sound intensity vector refined by {DNN}-based denoising
  and source separation,'' in \emph{Proc. {IEEE} Int. Conf. Acoust., Speech
  Signal Process.}, 2020, pp. 651--655.

\bibitem{pak2019sound}
J.~Pak and J.~W. Shin, ``Sound localization based on phase difference
  enhancement using deep neural networks,'' \emph{IEEE/ACM Trans. Audio,
  Speech, Lang. Process.}, vol.~27, no.~8, pp. 1335--1345, 2019.

\bibitem{cui2015data}
X.~Cui, V.~Goel, and B.~Kingsbury, ``Data augmentation for deep neural network
  acoustic modeling,'' \emph{IEEE/ACM Trans. Audio, Speech, Lang. Process.},
  vol.~23, no.~9, pp. 1469--1477, 2015.

\bibitem{salamon2017deep}
J.~Salamon and J.~P. Bello, ``Deep convolutional neural networks and data
  augmentation for environmental sound classification,'' \emph{IEEE Signal
  Process. Lett.}, vol.~24, no.~3, pp. 279--283, 2017.

\bibitem{perez2017effectiveness}
L.~Perez and J.~Wang, ``The effectiveness of data augmentation in image
  classification using deep learning,'' \emph{arXiv preprint:1712.04621}, 2017.

\bibitem{simard2003best}
P.~Y. Simard, D.~Steinkraus, J.~C. Platt \emph{et~al.}, ``Best practices for
  convolutional neural networks applied to visual document analysis,'' in
  \emph{Proc. Int. Conf. Docum. Anal. Recognit.}, 2003, pp. 958--963.

\bibitem{politis2020dataset}
A.~Politis, S.~Adavanne, and T.~Virtanen, ``A dataset of reverberant spatial
  sound scenes with moving sources for sound event localization and
  detection,'' in \emph{Proc. Detection Classification Acoust. Scenes Events
  Workshop}, Tokyo, Japan, November 2020, pp. 165--169.

\bibitem{pertila2021mobile}
P.~Pertil{\"a}, E.~Cakir, A.~Hakala, E.~Fagerlund, T.~Virtanen, A.~Politis, and
  A.~Eronen, ``Mobile microphone array speech detection and localization in
  diverse everyday environments,'' in \emph{Proc. {IEEE} Eur. Signal Process.
  Conf.}, 2021, pp. 406--410.

\bibitem{brousmiche2020secl}
M.~Brousmiche, J.~Rouat, and S.~Dupont, ``{SECL-UM}ons database for sound event
  classification and localization,'' in \emph{Proc. {IEEE} Int. Conf. Acoust.,
  Speech Signal Process.}, 2020, pp. 756--760.

\bibitem{nagatomo2022wearable}
K.~Nagatomo, M.~Yasuda, K.~Yatabe, S.~Saito, and Y.~Oikawa, ``Wearable {SELD}
  dataset: Dataset for sound event localization and detection using wearable
  devices around head,'' in \emph{Proc. {IEEE} Int. Conf. Acoust., Speech
  Signal Process.}, 2022, pp. 156--160.

\bibitem{evers2020locata}
C.~Evers, H.~W. L{\"o}llmann, H.~Mellmann, A.~Schmidt, H.~Barfuss, P.~A.
  Naylor, and W.~Kellermann, ``The {LOCATA C}hallenge: Acoustic source
  localization and tracking,'' \emph{IEEE/ACM Trans. Audio, Speech, Lang.
  Process.}, vol.~28, pp. 1620--1643, 2020.

\bibitem{politis2021dataset}
A.~Politis, S.~Adavanne, D.~Krause, A.~Deleforge, P.~Srivastava, and
  T.~Virtanen, ``A dataset of dynamic reverberant sound scenes with directional
  interferers for sound event localization and detection,'' in \emph{Proc.
  Detection Classification Acoust. Scenes Events Workshop}, Barcelona, Spain,
  November 2021, pp. 125--129.

\bibitem{guizzo2022l3das22}
E.~Guizzo, C.~Marinoni, M.~Pennese, X.~Ren, X.~Zheng, C.~Zhang, B.~Masiero,
  A.~Uncini, and D.~Comminiello, ``{L3DAS22 C}hallenge: Learning {3D} audio
  sources in a real office environment,'' in \emph{Proc. {IEEE} Int. Conf.
  Acoust., Speech Signal Process.}, 2022, pp. 9186--9190.

\bibitem{takahashi2017aenet}
N.~Takahashi, M.~Gygli, and L.~Van~Gool, ``Aenet: Learning deep audio features
  for video analysis,'' \emph{IEEE Trans. Multimedia}, vol.~20, no.~3, pp.
  513--524, 2017.

\bibitem{zhang2017mixup}
H.~Zhang, M.~Cisse, Y.~N. Dauphin, and D.~Lopez-Paz, ``mixup: Beyond empirical
  risk minimization,'' \emph{arXiv preprint arXiv:1710.09412}, 2017.

\bibitem{lu2017bidirectional}
R.~Lu and Z.~Duan, ``Bidirectional {GRU} for sound event detection,'' DCASE2017
  Challenge, Tech. Rep., September 2017.

\bibitem{takahashi2016deep}
N.~Takahashi, M.~Gygli, B.~Pfister, and L.~Van~Gool, ``Deep convolutional
  neural networks and data augmentation for acoustic event detection,''
  \emph{arXiv preprint arXiv:1604.07160}, 2016.

\bibitem{shimada2020sound}
K.~Shimada, N.~Takahashi, S.~Takahashi, and Y.~Mitsufuji, ``Sound event
  localization and detection using activity-coupled cartesian {DOA} vector and
  {RD3N}et,'' \emph{arXiv preprint arXiv:2006.12014}, 2020.

\bibitem{he2021neural}
W.~He, P.~Motlicek, and J.-M. Odobez, ``Neural network adaptation and data
  augmentation for multi-speaker direction-of-arrival estimation,''
  \emph{IEEE/ACM Trans. Audio, Speech, Lang. Process.}, vol.~29, pp.
  1303--1317, 2021.

\bibitem{mazzon2019first}
L.~Mazzon, Y.~Koizumi, M.~Yasuda, and N.~Harada, ``First order ambisonics
  domain spatial augmentation for {DNN}-based direction of arrival
  estimation,'' in \emph{Proc. Detection Classification Acoust. Scenes Events
  Workshop}, 2019, pp. 154--158.

\bibitem{higuchi2016robust}
T.~Higuchi, N.~Ito, T.~Yoshioka, and T.~Nakatani, ``Robust {MVDR} beamforming
  using time-frequency masks for online/offline {ASR} in noise,'' in
  \emph{Proc. {IEEE} Int. Conf. Acoust., Speech Signal Process.}, 2016, pp.
  5210--5214.

\bibitem{warsitz2007blind}
E.~Warsitz and R.~Haeb-Umbach, ``Blind acoustic beamforming based on
  generalized eigenvalue decomposition,'' \emph{IEEE/ACM Trans. Audio, Speech,
  Lang. Process.}, vol.~15, no.~5, pp. 1529--1539, 2007.

\bibitem{park2019specaugment}
D.~S. Park, W.~Chan, Y.~Zhang, C.-C. Chiu, B.~Zoph, E.~D. Cubuk, and Q.~V. Le,
  ``Spec{A}ugment: A simple data augmentation method for automatic speech
  recognition,'' \emph{arXiv preprint arXiv:1904.08779}, 2019.

\bibitem{zhang2019data}
J.~Zhang, W.~Ding, and L.~He, ``Data augmentation and prior knowledge-based
  regularization for sound event localization and detection,'' DCASE2019
  Challenge, Tech. Rep., June 2019.

\bibitem{Du2020_task3_report}
Q.~Wang, H.~Wu, Z.~Jing, F.~Ma, Y.~Fang, Y.~Wang, T.~Chen, J.~Pan, J.~Du, and
  C.-H. Lee, ``The {USTC}-i{F}lytek system for sound event localization and
  detection of {DCASE}2020 challenge,'' DCASE2020 Challenge, Tech. Rep., July
  2020.

\bibitem{DCASE2020_task3_report}
\BIBentryALTinterwordspacing
DCASE2020, ``Sound event localization and detection challenge results,''
  DCASE2020 Challenge, Tech. Rep., July 2020. [Online]. Available:
  \url{https://dcase.community/challenge2020/task-sound-event-localization-and-detection-results}
\BIBentrySTDinterwordspacing

\bibitem{gulati2020conformer}
A.~Gulati, J.~Qin, C.-C. Chiu, N.~Parmar, Y.~Zhang, J.~Yu, W.~Han, S.~Wang,
  Z.~Zhang, Y.~Wu \emph{et~al.}, ``Conformer: Convolution-augmented transformer
  for speech recognition,'' \emph{arXiv preprint arXiv:2005.08100}, 2020.

\bibitem{Du_NERCSLIP_task3_report}
Q.~Wang, L.~Chai, H.~Wu, Z.~Nian, S.~Niu, S.~Zheng, Y.~Wang, L.~Sun, Y.~Fang,
  J.~Pan, J.~Du, and C.-H. Lee, ``The {NERC-SLIP} system for sound event
  localization and detection of {DCASE}2022 challenge,'' DCASE2022 Challenge,
  Tech. Rep., June 2022.

\bibitem{DCASE2022_task3_report}
\BIBentryALTinterwordspacing
DCASE2022, ``Sound event localization and detection challenge results,''
  DCASE2022 Challenge, Tech. Rep., June 2022. [Online]. Available:
  \url{https://dcase.community/challenge2022/task-sound-event-localization-and-detection-evaluated-in-real-spatial-sound-scenes-results}
\BIBentrySTDinterwordspacing

\bibitem{kurz2015comparison}
E.~Kurz, F.~Pfahler, and M.~Frank, ``Comparison of first-order {A}mbisonics
  microphone arrays,'' in \emph{Proc. Int. Conf. Spatial Audio}, 2015.

\bibitem{politis2017comparing}
A.~Politis and H.~Gamper, ``Comparing modeled and measurement-based spherical
  harmonic encoding filters for spherical microphone arrays,'' in \emph{Proc.
  IEEE Workshop Appl. Signal Process. Audio Acoust.}, 2017, pp. 224--228.

\bibitem{Adavanne2019_DCASE}
S.~Adavanne, A.~Politis, and T.~Virtanen, ``A multi-room reverberant dataset
  for sound event localization and detection,'' in \emph{Proc. Detection
  Classification Acoust. Scenes Events Workshop}, New York University, NY, USA,
  October 2019, pp. 10--14.

\bibitem{politis2022starss22}
A.~Politis, K.~Shimada, P.~Sudarsanam, S.~Adavanne, D.~Krause, Y.~Koyama,
  N.~Takahashi, S.~Takahashi, Y.~Mitsufuji, and T.~Virtanen, ``{STARSS22}: A
  dataset of spatial recordings of real scenes with spatiotemporal annotations
  of sound events,'' \emph{arXiv preprint arXiv:2206.01948}, 2022.

\bibitem{guizzo2021l3das21}
E.~Guizzo, R.~F. Gramaccioni, S.~Jamili, C.~Marinoni, E.~Massaro, C.~Medaglia,
  G.~Nachira, L.~Nucciarelli, L.~Paglialunga, M.~Pennese \emph{et~al.},
  ``{L3DAS21 C}hallenge: Machine learning for {3D} audio signal processing,''
  in \emph{Proc. {IEEE} Int. Workshop Mach. Learning Signal Process.}, 2021,
  pp. 1--6.

\bibitem{acoustics2013em32}
\BIBentryALTinterwordspacing
mh~acoustics, ``{EM}32 {E}igenmike microphone array release notes (v17. 0),''
  Oct. 2013. [Online]. Available:
  \url{www.mhacoustics.com/sites/default/files/ReleaseNotes.pdf}
\BIBentrySTDinterwordspacing

\bibitem{daniel2000representation}
\BIBentryALTinterwordspacing
J.~Daniel, ``Repr{\'e}sentation de champs acoustiques, application {\`a} la
  transmission et {\`a} la reproduction de sc{\`e}nes sonores complexes dans un
  contexte multim{\'e}dia,'' Ph.D. dissertation, Univ. of Paris VI, France,
  2000. [Online]. Available: \url{http://gyronymo. free. fr}
\BIBentrySTDinterwordspacing

\bibitem{MazzonYasuda2019}
L.~Mazzon, M.~Yasuda, Y.~Koizumi, and N.~Harada, ``Sound event localization and
  detection using foa domain spatial augmentation,'' DCASE2019 Challenge, Tech.
  Rep., June 2019.

\bibitem{duong2010under}
N.~Q. Duong, E.~Vincent, and R.~Gribonval, ``Under-determined reverberant audio
  source separation using a full-rank spatial covariance model,'' \emph{IEEE
  Trans. Audio, Speech, Lang. Process.}, vol.~18, no.~7, pp. 1830--1840, 2010.

\bibitem{wang2021model}
Q.~Wang, H.~Wu, Z.~Jing, F.~Ma, Y.~Fang, Y.~Wang, T.~Chen, J.~Pan, J.~Du, and
  C.-H. Lee, ``A model ensemble approach for sound event localization and
  detection,'' in \emph{Proc. IEEE Int. Symposium Chinese Spoken Lang.
  Process.}, 2021, pp. 1--5.

\bibitem{he2016deep}
K.~He, X.~Zhang, S.~Ren, and J.~Sun, ``Deep residual learning for image
  recognition,'' in \emph{Proc. IEEE Conf. Comput. Vision Pattern Recogit.},
  2016, pp. 770--778.

\bibitem{chollet2017xception}
F.~Chollet, ``Xception: Deep learning with depthwise separable convolutions,''
  in \emph{Proc. IEEE Conf. Comput. Vision Pattern Recogit.}, 2017, pp.
  1251--1258.

\bibitem{povey2018semi}
D.~Povey, G.~Cheng, Y.~Wang, K.~Li, H.~Xu, M.~Yarmohammadi, and S.~Khudanpur,
  ``Semi-orthogonal low-rank matrix factorization for deep neural networks.''
  in \emph{Proc. Interspeech}, 2018, pp. 3743--3747.

\bibitem{vaswani2017attention}
A.~Vaswani, N.~Shazeer, N.~Parmar, J.~Uszkoreit, L.~Jones, A.~N. Gomez,
  {\L}.~Kaiser, and I.~Polosukhin, ``Attention is all you need,'' in \emph{Adv.
  Neural Inf. Process. Syst.}, 2017, pp. 5998--6008.

\bibitem{chen2020continuous}
S.~Chen, Y.~Wu, Z.~Chen, J.~Li, C.~Wang, S.~Liu, and M.~Zhou, ``Continuous
  speech separation with conformer,'' \emph{arXiv preprint arXiv:2008.05773},
  2020.

\bibitem{Miyazaki2020}
K.~Miyazaki, T.~Komatsu, T.~Hayashi, S.~Watanabe, T.~Toda, and K.~Takeda,
  ``Convolution-augmented transformer for semi-supervised sound event
  detection,'' DCASE2020 Challenge, Tech. Rep., June 2020.

\bibitem{Cao2019}
Y.~Cao, T.~Iqbal, Q.~Kong, M.~Galindo, W.~Wang, and M.~Plumbley, ``Two-stage
  sound event localization and detection using intensity vector and generalized
  cross-correlation,'' DCASE2019 Challenge, Tech. Rep., June 2019.

\bibitem{mesaros2019joint}
A.~Mesaros, S.~Adavanne, A.~Politis, T.~Heittola, and T.~Virtanen, ``Joint
  measurement of localization and detection of sound events,'' in \emph{Proc.
  IEEE Workshop Appl. Signal Process. Audio Acoust.}, 2019, pp. 333--337.

\bibitem{kingma2014adam}
D.~P. Kingma and J.~Ba, ``Adam: A method for stochastic optimization,''
  \emph{arXiv preprint arXiv:1412.6980}, 2014.

\end{thebibliography}

%



%
\begin{IEEEbiography}[{\includegraphics[width=1in,height=1.25in,clip,keepaspectratio]{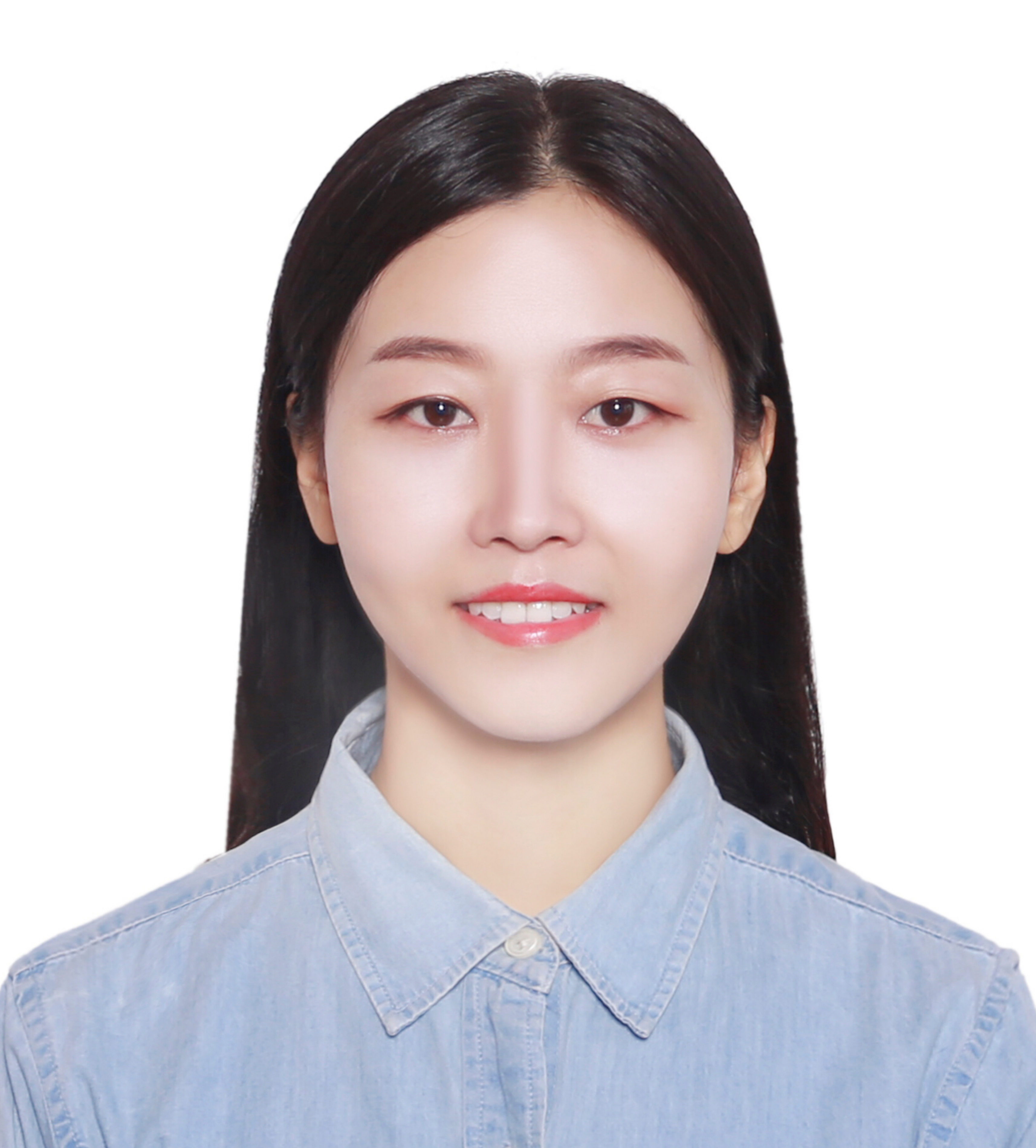}}]{Qing Wang}
received the B.S. and Ph.D. degrees from the Department of Electronic Engineering and Information Science, University of Science and Technology of China (USTC), Hefei, China, in 2012 and 2018, respectively. From July 2018 to February 2020, she worked at Tencent company on speech enhancement. From March 2020 to February 2023, she was a Postdoctor at USTC. She is currently an Assistant Professor at USTC. Her research interests include speech enhancement, acoustic scene classification, sound event localization and detection.
\end{IEEEbiography}

\begin{IEEEbiography}[{\includegraphics[width=1in,height=1.25in,clip,keepaspectratio]{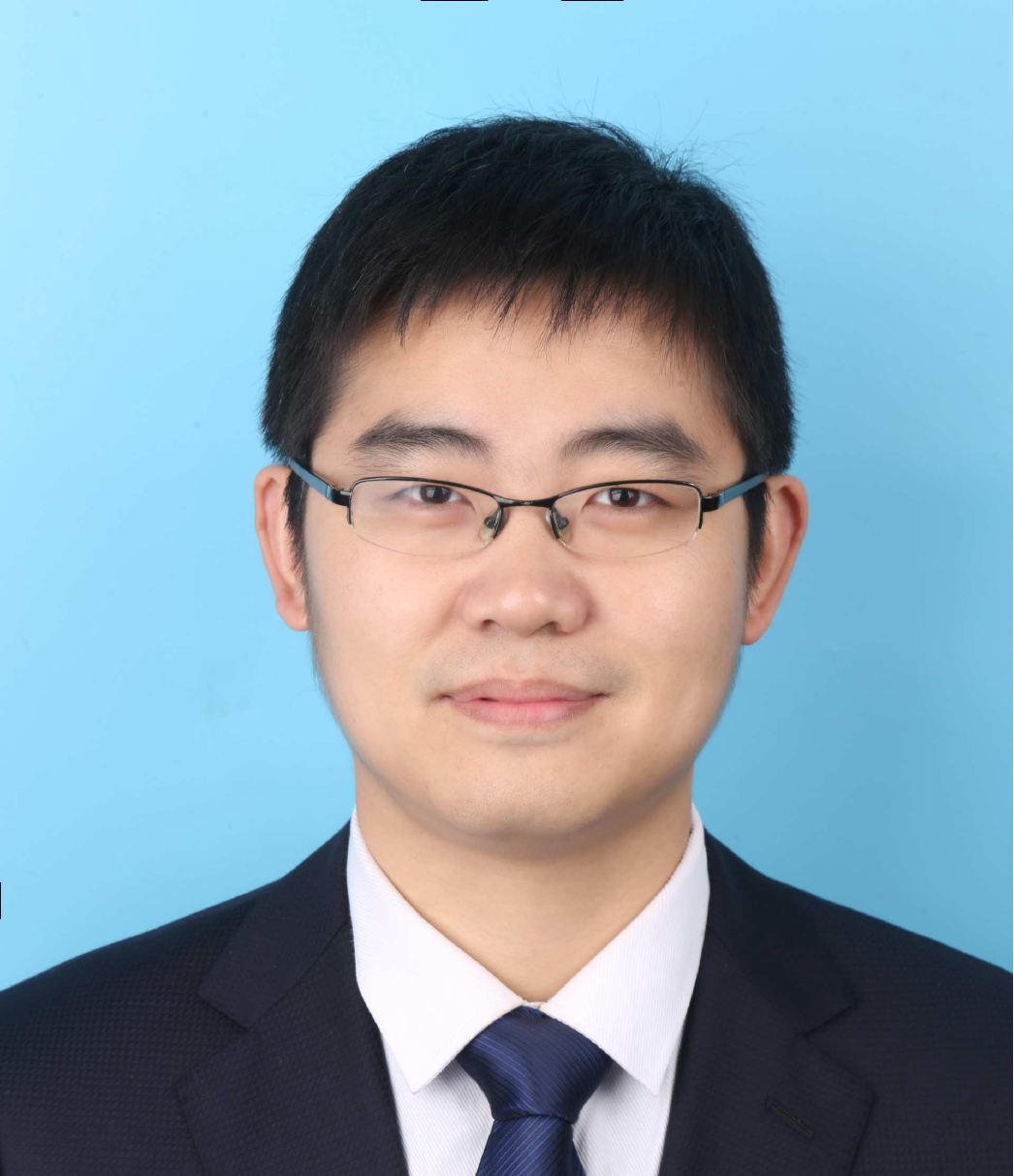}}]{Jun Du}
(Senior Member, IEEE) received the B.Eng. and Ph.D. degrees from the Department of Electronic Engineering and Information Science, University of Science and Technology of China (USTC), Hefei, China, in 2004 and 2009, respectively. From 2009 to 2010, he was with iFlytek Research as a Team Leader, working on speech recognition. From 2010 to 2013, he joined Microsoft Research Asia as an Associate Researcher, working on handwriting recognition, OCR. Since 2013, he has been with the National Engineering Laboratory for Speech and Language Information Processing, USTC. He has authored or coauthored more than 150 papers. His main research interests include speech signal processing and pattern recognition applications. He is an Associate Editor for the IEEE/ACM TRANSACTIONS ON AUDIO, SPEECH AND LANGUAGE PROCESSING and a Member of the IEEE Speech and Language Processing Technical Committee. He was the recipient of the 2018 IEEE Signal Processing Society Best Paper Award. His team won several champions of CHiME-4/CHiME-5/CHiME-6 Challenge, SELD Task of 2020 DCASE Challenge, and DIHARD-III Challenge.
\end{IEEEbiography}

\begin{IEEEbiography}[{\includegraphics[width=1in,height=1.25in,clip,keepaspectratio]{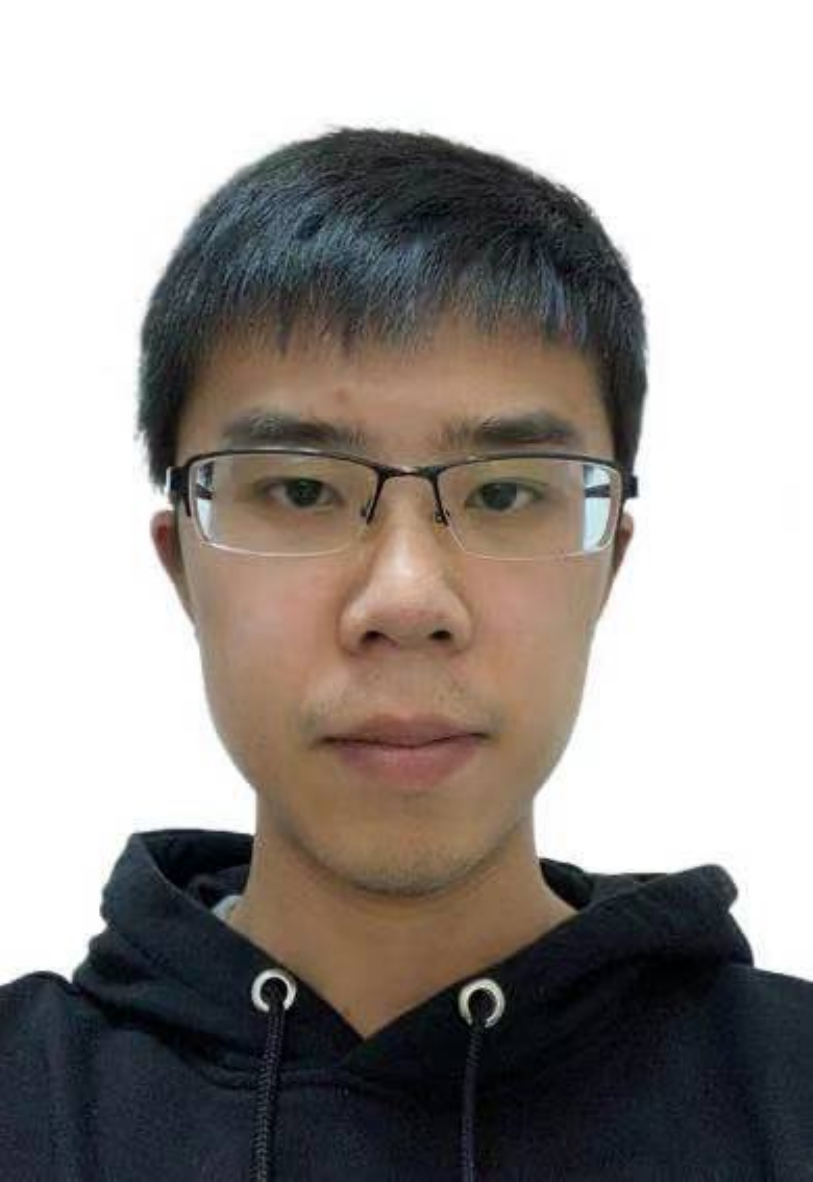}}]{Hua-Xin Wu}
received the B.E. degrees in 2016 from the Southeast University. Since 2016, he has been with iFlytek Research on multimodal speech recognition and keyword spotting. His current research interests include keyword spotting and sound event detection.
\end{IEEEbiography}

\begin{IEEEbiography}[{\includegraphics[width=1in,height=1.25in,clip,keepaspectratio]{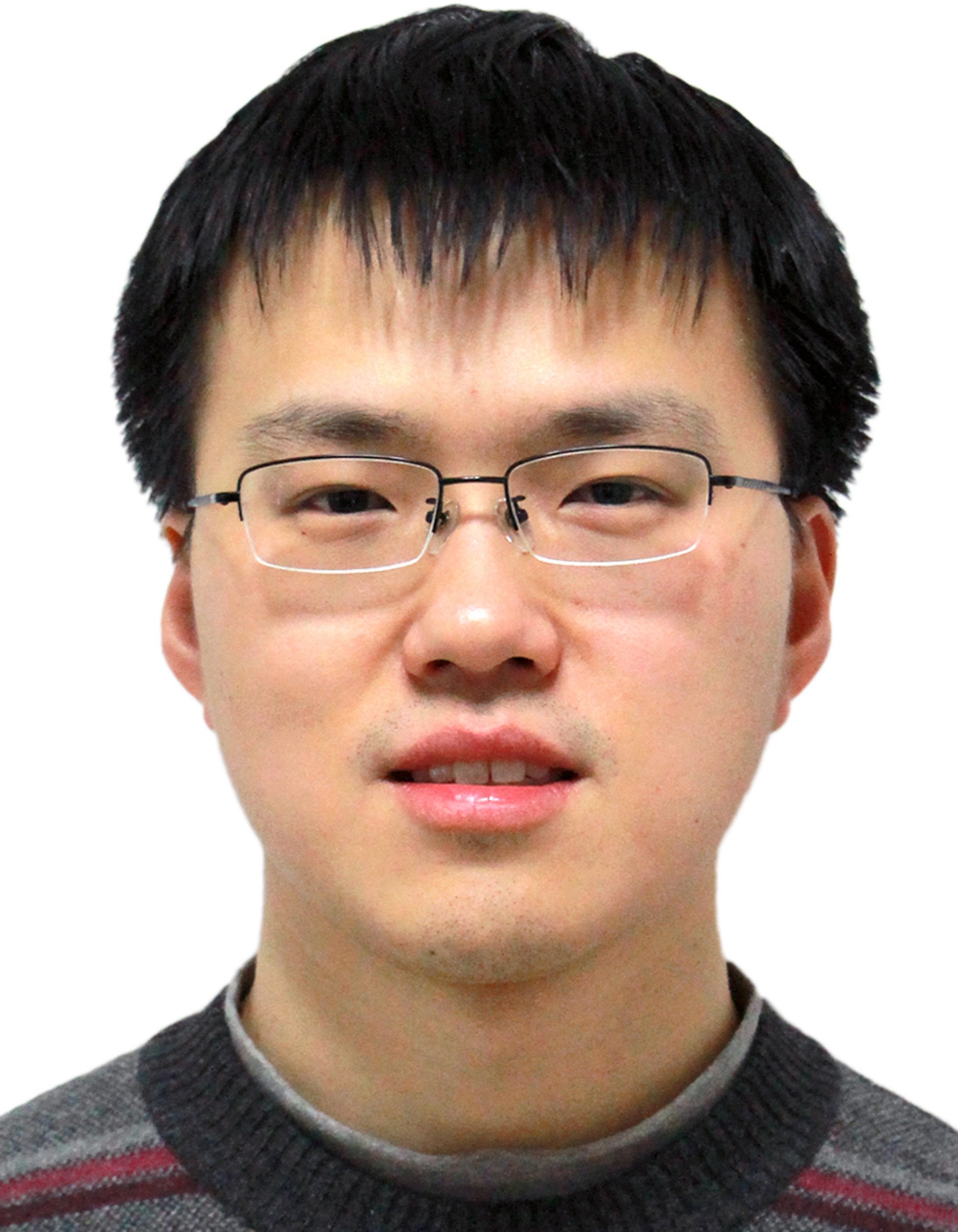}}]{Jia Pan}
received the B.S., M.S. and Ph.D. degrees in 2006, 2009 and 2020, respectively, from the Department of Electronic Engineering and Information Science, University of Science and Technology of China, Hefei, China. Since 2009, he has been with iFlytek Research on speech recognition and spoken dialogue systems. His current research interests include speech recognition and machine learning.
\end{IEEEbiography}

\begin{IEEEbiography}[{\includegraphics[width=1in,height=1.25in,clip,keepaspectratio]{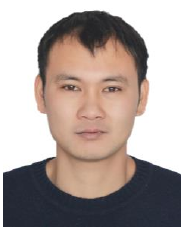}}]{Feng Ma}
received the B.Eng. and M.S. degrees from the Department of Electronic Engineering and Information Science, University of Science and Technology of China, Hefei, China, in 2009 and 2012, respectively. He is currently with iFlytek Research, Hefei, China. His current research interests include acoustic echo cancellation, microphone arrays, and robust speech recognition.
\end{IEEEbiography}

\begin{IEEEbiography}[{\includegraphics[width=1in,height=1.25in,clip,keepaspectratio]{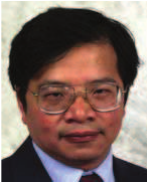}}]{Chin-Hui Lee}
(Fellow, IEEE) is a Professor with the School of Electrical and Computer Engineering, Georgia Institute of Technology. Before joining academia in 2001, he had 20 years of industrial experience, ending at Bell Laboratories, Murray Hill, NJ, USA, as a Distinguished Member of Technical Staff, and the Director of the Dialogue Systems Research Department. He has authored or coauthored more than 550 papers and 30 patents, and has been cited more than 50 000 times for his original contributions with an h-index of 80 on Google Scholar. He has received numerous awards, including the Bell Labs President's Gold Award in 1998. He also won SPS's 2006 Technical Achievement Award for ``Exceptional Contributions to the Field of Automatic Speech Recognition''. In 2012, he was invited by ICASSP to give a plenary talk on the future of speech recognition. In the same year, he was awarded the ISCA Medal in scientific achievement for pioneering and seminal contributions to the principles and practice of automatic speech and speaker recognition. He is a Fellow of ISCA.
\end{IEEEbiography}




\end{document}